\documentclass{article}

\usepackage{PRIMEarxiv}

\usepackage[utf8]{inputenc} 
\usepackage[T1]{fontenc}    
\usepackage{hyperref}       
\usepackage{url}            
\usepackage{booktabs}       
\usepackage{amsfonts}       
\usepackage{nicefrac}       
\usepackage{microtype}      
\usepackage{lipsum}
\usepackage{fancyhdr}       
\usepackage{graphicx}       
\graphicspath{{media/}}     

\newcommand{\teightfifty}{\ensuremath{T_{850}}}
\newcommand{\dteightfifty}{\ensuremath{\Delta T_{850}}}
\newcommand{\latentvector}{\ensuremath{\mathbf{z}}}

\newcommand{\deterministicmultiplier}{\ensuremath{A}}

\newcommand{\latentB}{\ensuremath{\mathbf{B}_{z}}}

\usepackage{siunitx}
\usepackage{dsfont}
\usepackage{subcaption}
\usepackage{lineno}
\usepackage{bm}
\usepackage{amsmath}
\usepackage{natbib}
\usepackage{authblk}

\title{3D-Var Data Assimilation using a Variational Autoencoder
\thanks{\textit{\underline{Peer-reviewed version published in QJRMS (open access)}}: 
\href{https://doi.org/10.1002/qj.4708}{https://doi.org/10.1002/qj.4708}} 
}

\author[ ]{%
  \begin{minipage}{0.4\textwidth}
    \centering
    \textbf{Bo\v{s}tjan Melinc}\textsuperscript{1} \\
    \texttt{bostjan.melinc@fmf.uni-lj.si}
  \end{minipage}%
  \vspace{3mm}
}
\author[ ]{%
  \begin{minipage}{0.4\textwidth}
    \centering
    \textbf{\v{Z}iga Zaplotnik}\textsuperscript{2,1} \\
    \texttt{ziga.zaplotnik@ecmwf.int}
  \end{minipage}%
  \vspace{3mm}
}

\affil[1]{University of Ljubljana, Faculty of Mathematics and Physics, \protect\\Jadranska 19, 1000 Ljubljana, Slovenia\vspace{4mm}}
\affil[2]{ECMWF, \protect\\ Shinfield Park, Reading, RG2 9AX Code, United Kingdom}

\date{} 

\begin{document}
\maketitle

\begin{abstract}
Data assimilation of atmospheric observations traditionally relies on variational and Kalman filter methods. Here, an alternative neural-network data assimilation (NNDA) with variational autoencoder (VAE) is proposed. The three-dimensional variational (3D-Var) data assimilation cost function is utilised to determine the analysis that optimally fuses simulated observations and the encoded short-range persistence forecast (background), accounting for their errors. The minimisation is performed in the reduced-order latent space, discovered by the VAE.  The variational problem is auto-differentiable, simplifying the computation of the cost function gradient necessary for efficient minimisation. 
We demonstrate that the background-error covariance ($\mathbf{B}$) matrix measured and represented in the latent space is quasi-diagonal. The background-error covariances in the grid-point space are flow-dependent, evolving seasonally and depending on the current state of the atmosphere. Data assimilation experiments with a single temperature observation in the lower troposphere indicate that the $\mathbf{B}$-matrix simultaneously describes both tropical and extratropical background-error covariances.
\end{abstract}


\section{Introduction}
In recent years, significant progress has been made through the use of machine learning in weather prediction. These developments encompass various aspects of the numerical weather prediction (NWP) workflow, including the use of neural-network (NN) emulators for specific components, such as radiation schemes \citep[e.g.][]{Meyer2022}, convection schemes \citep[e.g.][]{Yuval2020}, bias-correction and forecast postprocessing \citep[e.g.][]{Kim2021,Frnda2022}, uncertainty estimation \citep[e.g.][]{Clare2021} and ensemble spread estimation \citep[e.g.][]{Brecht2023a}. The availability of ERA5 reanalyses \citep{Hersbach2020} through Copernicus Climate Data Store boosted an immense advance in full NN weather prediction models \citep[e.g.][]{Keisler2022,Pathak2022a,Bi2023a,Lam2023,Nguyen2023,Chen2023fengwu,Chen2023a}. Most of these weather prediction tools have demonstrated comparable large-scale forecast skill at medium-range lead times to the high-resolution deterministic forecasts of the world-leading NWP system, the Integrated Forecast System (IFS) of the European Centre for Medium-range Weather Forecasts (ECMWF). It is important to note that these NN models decently simulate the atmospheric dynamics \citep{Hakim2023}, but fail to simulate error growth \citep{SelzCraig2023}, and suffer from field smoothing and vanishing precipitation and power spectra with increasing forecasts lead time \citep{Bonavita2023, Bouallegue2023, Rasp2023}. Consequently, they cannot be considered as true weather emulators, and should be more fairly compared against the ensemble mean of the ensemble prediction system (EPS) \citep[as in][]{Chen2023_fuxi}. Moreover, these models inevitably inherit the biases from the ERA5 ground truth on which they are trained. 

Another limitation of pure NN models is their inability to independently produce an operational forecast, as they rely on initial conditions provided by the operational NWP centers through the data assimilation process. Data assimilation is a methodology that optimally fuses information from millions of recent Earth-system observations and short-range model simulations \citep{Kalnay2003,Lahoz2014} to  obtain the most accurate estimate of the current state of the Earth-system, known as the \textit{the analysis}, which serves as the initial state for the weather forecasts. 

The main motivation for such merging approach is that the observations are accurate, but also relatively sparse and typically biased. They do not cover the entire globe or capture all vertical slices of the atmosphere at each time instant. Furthermore, not all atmospheric, land, or ocean characteristics can be directly observed. For instance, instead of directly measuring profiles of temperature, humidity, clouds, and precipitation, satellites measure radiances that provide implicit information on these variables \citep[e.g.][]{Geer2018}. While the observations have irregular sampling, the forecast models provide a complete representation of the Earth-system state. However, forecast models are often less accurate than observations as the forecast error grows with time. Therefore, the observations and the short-range model forecast are objectively merged based on Bayesian inference by considering their respective error statistics. 

Neural networks have been applied to a somewhat lesser extent in data assimilation than in the forecasting part of NWP workflow, despite (or because of) significant mathematical similarities between machine learning and data assimilation \citep{Geer2021,Cheng2023a}. So far, NNs have been mostly used for specific tasks of the NWP data assimilation workflow. For instance, studies by \citet{Bonavita2020} and \citet{Laloyaux2022} have demonstrated that the NNs are similarly effective in estimating model biases as the weak-constraint formulation of 4D-Var data assimilation. NNs were also used to derive tangent linear model and adjoint model used in 4D-Var data assimilation \citep{Hatfield2021}, leveraging the advantageous property of easy auto-differentiation in NNs, a characteristic that is also exploited in this study. 

A full NN data assimilation in an NWP setting was recently demonstrated by \citet{deAlmeida2022}. They employed a dense NN and trained it to emulate analysis increments in Weather Research and Forecasting (WRF) model derived from the 3D-Var data assimilation of surface observations and atmospheric sounding data. In another study, \citet{Andrychowicz2023} performed \textit{implicit} data assimilation. They trained the convolutional-vision transformer NN model (MetNet-3) to predict the surface variables (temperature, dew point, wind speed and direction) up to 24 hours ahead by minimizing the difference between the model forecast and real surface observations. The training data used for forecast initialization included ground-based weather stations, GOES satellite observations, radar precipitation as well as the analysis from the High Resolution Rapid Refresh (HRRR) NWP model. The latter provided a complete initial representation of the state of the atmosphere (first guess), which is then further corrected by the observations. Their approach reaffirms the necessity to combine observations with a complete prior information of the earth-system state even in case of neural-network data assimilation (NNDA). NNDA has also been performed in a simplified model setting using augmented approach to determine both the model state and the model parameters \citep{Bocquet2020,Brajard2020,Fablet2021,Legler2022} or to correct the model dynamics \citep{Farchi2021}. \citet{Arcucci2021} has proposed an alternative merging approach of DA and NNs by using DA to iteratively correct the NN parameters based on upcoming observations.

In this study, we adhere to Bayesian fundamentals and employ a variational autoencoder (VAE) to demonstrate NNDA of simulated temperature observations at 850-hPa pressure level within a reduced-order latent space. Our method resembles the three-dimensional variational (3D-Var) data assimilation. While this represents the first employment the variational approach for neural-network latent space DA in NWP, prior attempts of variational DA have been conducted in reduced order latent spaces using linear empirical orthogonal functions \citep{Robert2005} or singular vectors \citep{Chai2007,Cheng2010}. Latent space DA has also been applied using Kalman filter methods with the Lorenz 96 model \citep{Peyron2021}, air quality models \citep{Amendola2021}, and in some other applications \citep[e.g.][]{Canchumuni2019,Zhuang2022}. The theoretical algorithm on variational latent space DA (although in a slightly different form) with a standard neural-network autoencoder has been described by \citet{Mack2020}.

The article is structured as follows. Section~\ref{section:da_vae} outlines the methodology of 3D-Var data assimilation using VAE, along with the background-error estimation and minimisation algorithm. Section~\ref{section:results} presents the results of observing system simulation experiments (OSSEs) involving a single temperature observation at various locations, as well as an experiment with simulated global temperature observations. Discussion, conclusion and outlook are given in Section~\ref{section:conclusions}.

\section{Data assimilation with a variational autoencoder}\label{section:da_vae}

\subsection{Data} 
\label{sec:data}
The training and evaluation of the VAE utilised temperature data at the 850 hPa pressure level (\teightfifty) from the ERA5 reanalysis \citep{Hersbach2020}. Daily mean data were derived from hourly data on a regular latitude-longitude grid with 0.25$^\circ$ resolution. Additionally, the data were latitudinally regridded to exclude the poles. 

To generate a standardised input for neural network training, the data were normalised by subtracting the 1981-2010 climatological mean and dividing it by the climatological standard deviation for each grid point and day of the year. The data were split as follows: the period 1979-2014 was used for NN training, 2015-2018 for validation, and 2019-2022 for testing.

\subsection{Representation of global 2D atmospheric field with a variational autoencoder}
\label{sec:VAE introduction}

To perform NNDA, we first learn the representation of the $\teightfifty$ atmospheric field in the latent space using convolutional VAE. Let us explain the VAE by first describing a standard autoencoder (AE). AE is a type of neural network which is trained to recreate its input as close as possible. It consists of two main parts: (1) the encoder $E^{AE}$, which maps the input state $\mathbf{x}$ to a state $\mathbf{z}$ in the latent space of reduced dimension, i.e. $\mathbf{z}=E^{AE}(\mathbf{x})$, and (2) the decoder $D^{AE}$, which mirrors the encoder and transforms the latent state back to the original space, so that the output $\mathbf{x}'$ is similar to the input $\mathbf{x}$, i.e.  $\mathbf{x}'\approx D^{AE}(E^{AE}(\mathbf{x}))$. In the case of meteorological fields, the input and output fields are typically described in a grid point space. 

The VAE also includes the encoder and decoder, however, the latent state $\mathbf{z}$ is not a deterministic function of the input. In the VAE, $\mathbf{z}$ is instead randomly sampled from the multivariate normal distribution whose parameters (the mean and the variance) are deterministic functions of the input.
Consequently, the same input state results in different outputs that are perturbed analogues of the input. If the VAE is properly trained, the output states are expected to represent realisations of climatologically possible states.  

Due to the stochastic nature of VAE, the loss function $\mathcal{L}$ that is minimised during the training contains two parts \citep{kingma2019introduction}: the \textit{reconstruction loss} $\mathcal{L}^{\mathrm{rec}}$, and the \textit{regularisation loss} $\mathcal{L}^{\mathrm{reg}}$, so that the final loss function is 
\begin{equation}\label{eq:total_loss}
    \mathcal{L} = \mathcal{L}^{\mathrm{rec}} + \mathcal{L}^{\mathrm{reg}} \, .
\end{equation} 
$\mathcal{L}^{\mathrm{rec}}$ measures how well the VAE reconstructs the input data. Learning the reconstruction involves solving an optimisation problem, where the objective is to find optimal model parameters $\bm{\theta}$ that maximise the probability $p_{\bm{\theta}}(\mathbf{x}')$ of the output state $\mathbf{x}'$ matching the input data $\mathbf{x}$, given the distribution $p_{\bm{\theta}}(\mathbf{z})$ of the latent state $\mathbf{z}$, and a generative decoder model $p_{\bm{\theta}}(\mathbf{x}'|\mathbf{z})$ parameterised by $\bm{\theta}$, i.e.
\begin{equation}\label{eq:optimisation}
    \max_{\bm{\theta}} p_{\bm{\theta}}(\mathbf{x}') = \max_{\bm{\theta}} \int_{\mathbf{z}} p_{\bm{\theta}}(\mathbf{x}' | \mathbf{z}) p_{\bm{\theta}}(\mathbf{z}) d\mathbf{z} \, .
\end{equation}
To simplify the optimisation problem, a recognition encoder model $q_{\bm{\phi}}(\mathbf{z}|\mathbf{x})$ is introduced, with parameters $\bm{\phi}$ determined such that $q_{\bm{\phi}}(\mathbf{z}|\mathbf{x}) \approx p_{\bm{\theta}}(\mathbf{z}|\mathbf{x}) = p_{\bm{\theta}}(\mathbf{x} | \mathbf{z}) p_{\bm{\theta}}(\mathbf{z}) / p_{\bm{\theta}}(\mathbf{x})$. After some derivation and assuming a Gaussian multivariate distribution of $q_{\bm{\phi}}(\mathbf{z}|\mathbf{x})$ with diagonal covariance, i.e. $q_{\bm{\phi}}(\mathbf{z}|\mathbf{x}) = \mathcal{N}\left(\bm{\mu},\bm{\Sigma}\right)$, $\bm{\Sigma}=\mathrm{diag}\left(\bm{\sigma^2}\right)$, and $p_{\bm{\theta}}(\mathbf{z})\approx\mathcal{N}(\bm{0},\mathbf{I})$, the reconstruction loss can be written as \citep{kingma2019introduction}:
\begin{equation}
    \mathcal{L}^{\mathrm{rec}}(\mathbf{x}') = - \log p_{\bm{\theta}}(\mathbf{x}' | \mathbf{z}) \, . 
\end{equation}
For this term, we chose the Huber loss function, multiplied by the deterministic multiplier $\deterministicmultiplier>0$, i.e.
\begin{equation}
    \mathcal{L}^{\mathrm{rec}}(\mathbf{x}',\mathbf{x}) = \deterministicmultiplier \,\frac{1}{n}\sum_{i=1}^n L_\delta (x'_i, x_i); 
    \qquad
    L_\delta (\zeta, \xi) =
    \begin{cases}
			\frac{1}{2} (\zeta - \xi)^2, & \text{if}\, |\zeta - \xi| \leq \delta, \\
            \delta \left(|\zeta - \xi| - \frac{1}{2}\delta\right), & \text{otherwise},
	\end{cases}
\end{equation}
where $x'_i$ and $x$ are the elements of the state vector $\mathbf{x}'$ and input state vector $\mathbf{x}$ with size $n$, respectively, and $\delta=1$. The deterministic multiplier controls the balance of the two loss terms and therefore the amount of stochasticity. The second part of the loss function, $\mathcal{L}^{\mathrm{reg}}$, enforces a generation of the latent state with a desired ($p_{\bm{\theta}}(\mathbf{z})\approx\mathcal{N}(\bm{0},\mathbf{I})$) distribution \citep{Goodfellow2016,kingma2019introduction}:
\begin{equation}\label{eq:regularisation_loss}
   \mathcal{L}^{\mathrm{reg}} =  - \log{p_{\bm{\theta}}(\mathbf{z})} + \log{q_{\bm{\phi}}}(\mathbf{z} | \mathbf{x}) \, .
\end{equation}

The encoder $q_{\bm{\phi}}(\mathbf{z}|\mathbf{x})$ produces vectors of means $\bm{\mu}_{\phi}$ and log-variances $\log{\bm{\sigma}_{\phi}^2}$ in the last stack of neurons (Fig.~\ref{fig:VAE_structure}), representing the multivariate Gaussian distribution. Each latent vector element $z_i$ is then randomly sampled from this distribution \citep{kingma2022autoencoding} as 
\begin{equation}\label{eq:sampling}
    z_i = \mu_{\phi i} + \hat{z}_i \sigma_{\phi i}, \quad \hat{z}_i\sim\mathcal{N}(0,1) \, .
\end{equation}
Consequently, the size of the encoder's output layer is twice the size of the latent space vector. 
The latent vector $\latentvector$ then enters the decoder to produce a reconstructed vector $\mathbf{x}'$. 

During the minimisation of loss function (\ref{eq:total_loss}), the first term in (\ref{eq:regularisation_loss}) drives the encoder so that the final distribution of the sampled $\latentvector$ will resemble $p_{\bm{\theta}}(\mathbf{z})\approx\mathcal{N}(\bm{0},\mathbf{I})$. On the other hand, the second term in (\ref{eq:regularisation_loss}) serves to increase the elements of $\bm{\Sigma}$, thereby enhancing the stochasticity of the VAE.
A trained decoder can be used to generate instances $\mathbf{x}'$ from climatological distribution, $\mathbf{x}'\sim p_{clim}(\mathbf{x}')$, by sampling $\mathbf{z}\sim\mathcal{N}(\bm{0},\mathbf{I})$. For more technical details on the derivation and implementation of the VAE, the reader is referred to \citet{kingma2019introduction} and the references therein.

In contrast to AE, the extra regularisation terms in VAE's loss function ensure a smooth latent space as well as a smooth transition from the latent to the grid point space \citep{LiH2020, Grooms2021}. Consequently, when two latent states are in proximity, their decoded counterparts typically exhibit close resemblance. 
This property is essential in our approach, enabling us to: 1) generate new climatologically plausible fields by randomly sampling latent states from a standard normal distribution, 2) perform variational data assimilation, and 3) generate ensembles of fields from a single ensemble member as demonstrated in \citet{Grooms2021}. 

\subsubsection{VAE setup and training}
\begin{figure}[ht!]
    \centering
    \includegraphics[width=\textwidth, clip, trim={0px 0px 0px 0px}]{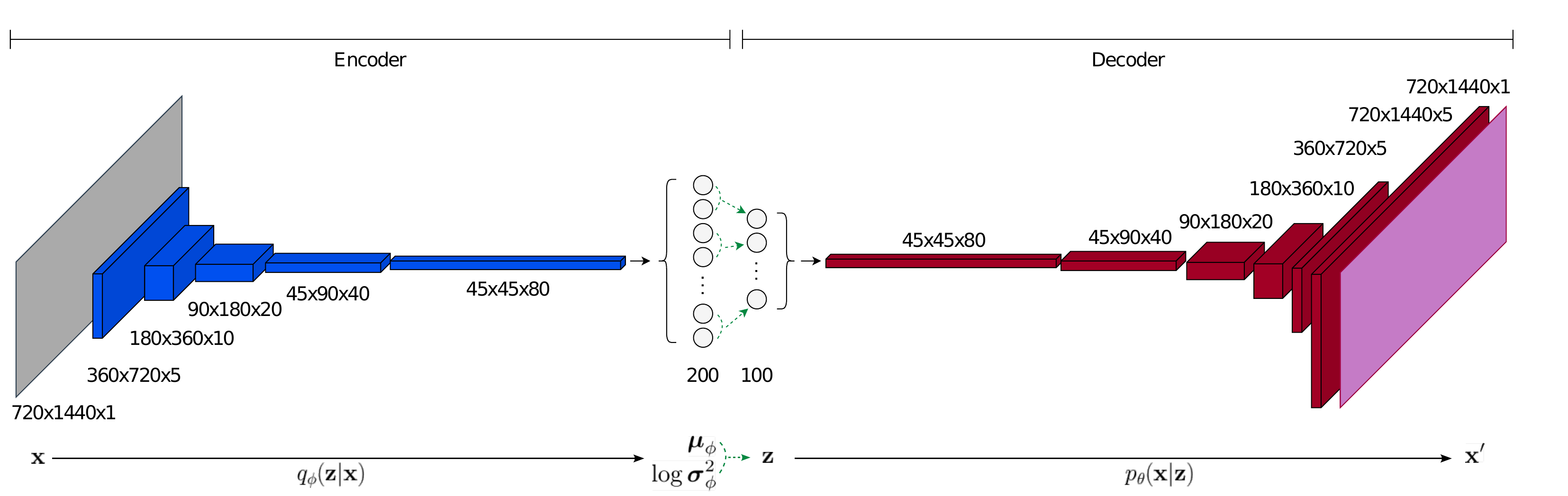}
    \caption{Structure of our VAE model. The input field (grey colour) with $720\times1440$ grid points in the meridional and zonal direction enters the encoder. Through a series of 2D convolutional layers, the size of the field gradually decreases while the number of channels increases. The intermediate fields (filters) are depicted in blue, with accompanying numbers indicating the feature size and the channel count. The final intermediate field then enters a dense layer with neurons (represented by grey circles) which denote the mean ($\mu_{\phi}$) and the logarithm of variance ($\log{\sigma_{\phi}^2}$) for each element of the latent vector $\mathbf{z}$. Each pair of parameters determines the Gaussian distribution of a single latent vector element, from which a value for a corresponding neuron in the decoder's input layer is sampled  (as indicated by the green dashed arrows). The input to the decoder is further transformed by another dense layer into a field with dimensions of $45\times45$ and 80 channels. The intermediate fields in the decoder are shown in red. As the fields pass through 2D transposed convolutional layers, their size gradually increases while the number of channels decreases. The output field from the decoder (pink colour) matches the shape of the input field to the encoder.}
    \label{fig:VAE_structure}
\end{figure}

The architecture of our VAE network is described in Figure~\ref{fig:VAE_structure} and follows \citet{Brohan_Machine_Learning_for_2022} with some additional fine-tuning of the training parameters and network architecture.
The size of the latent vector was $N=100$~elements. The encoder and decoder employed 2D convolutional layers and 2D transposed convolutional layers with $3\times3$ kernels and either $2\times2$ or $1\times2$ strides for field reshaping. In 2D convolutional layers, we applied periodic padding in the zonal direction and padding over each of the poles. The output layers of both the encoder and the decoder utilised the linear activation function. In the decoder's dense layer, a rectified linear unit (ReLU) activation function was used, while the exponential linear unit (ELU) activation function was applied elsewhere. As discussed in Section~\ref{sec:data}, the input fields of daily mean~$\teightfifty$ were standardised by subtracting the climatological mean of the day-of-year and dividing by the climatological standard deviation of the day-of-year.

We explored several configurations with different deterministic multipliers~$\deterministicmultiplier$ and trained the NN for 100 epochs per setup. We observed that a large $\deterministicmultiplier$ resulted in a relatively smaller Huber norm in comparison to smaller $\deterministicmultiplier$. However, the convergence rate of $p_{\bm{\theta}}(\latentvector)$ towards the standard normal distribution was much slower than the rate at which the overfitting to the training set was occurring. Conversely, a very small~$\deterministicmultiplier$ resulted in a minimal decrease in the Huber norm during training, making the VAE ineffective at reconstructing the input fields. 

We also assessed the global mean $\teightfifty$ anomaly of the reconstructed states, defined as the output $\teightfifty$ minus the climatological global mean temperature for the specific day of the year. For the optimal deterministic multiplier $\deterministicmultiplier$, the average global mean $\teightfifty$ anomaly, obtained by averaging 150 perturbed $\teightfifty$ states decoded from 150 randomly sampled ($\mathbf{z}\sim\mathcal{N}(\bm{0},\mathbf{I})$) latent states, was below $\SI{0.05}{\degreeCelsius}$ after approximately 10 epochs of training (ideally, it should be zero). The $p_{\bm{\theta}}(\latentvector)$ approached $\mathcal{N}(\bm{0},\mathbf{I})$ closely after 20 epochs, with convergence slowing thereafter. Although the loss function score for the validation set already reached its global minimum already at the 11th epoch, we opted to use the weights from the 20th epoch for our subsequent experiments. At this epoch, the network exhibited only a slight overfitting to the training set; the loss function value for the validation set increased by a mare 0.8\% compared to the 11th epoch.

\subsubsection{Properties of VAE output}

In order to emphasise the details of the temperature fields, we chose to visualise the temperature anomalies ($\dteightfifty$) instead of the raw temperature fields throughout the paper. Temperature anomalies are obtained by subtracting the climatological mean values from the destandardised output of the decoder. For simplicity, we will commonly refer to these anomalies as \textit{decoded fields}. When presenting these fields, we may even omit "decoded" as it is evident that the gridded fields are not in the latent space. In the continuation, we will denote the encoder of the input state $q_{\bm{\phi}}(\mathbf{z}|\mathbf{x})$ as $E(\mathbf{x})$ and the decoder of the latent state $p_{\bm{\theta}}(\mathbf{x}'|\mathbf{z})$ as $D(\mathbf{z})$.
When utilising only the unperturbed part of the encoder to generate  the latent state (i.e.~$\mathbf{z}=\boldsymbol{\mu_\phi}$), we will denote the encoder of the input state as $\mu_\phi(\mathbf{x})$.

\begin{figure}[h!]
    \centering
    \includegraphics[width=\textwidth]{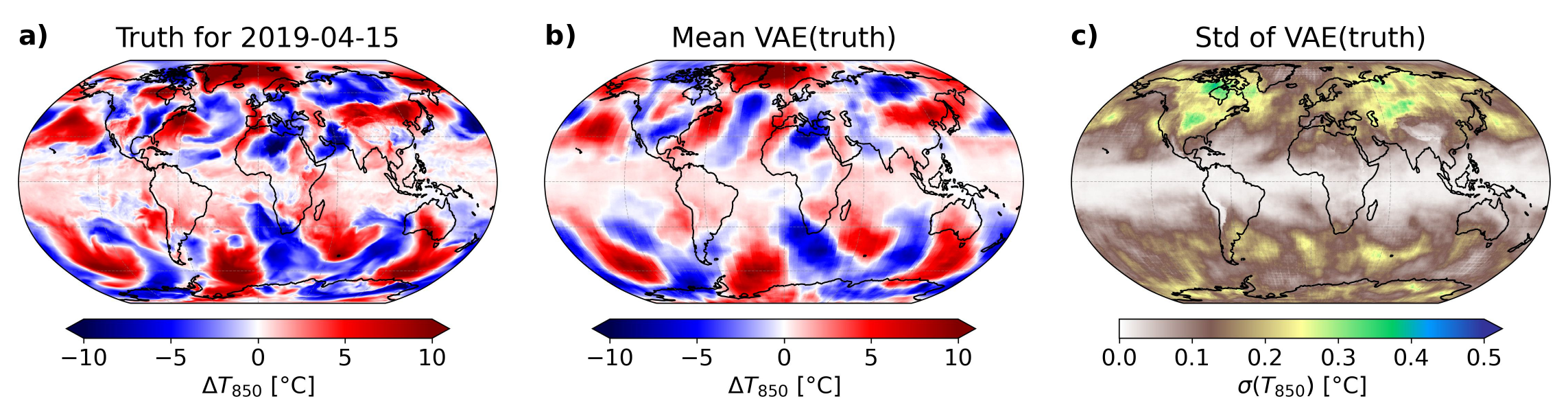}
    \caption{(a) The ground truth climatological temperature anomaly on April 15, 2019. (b) The ensemble mean of VAE output fields, generated by feeding the truth from (a) to the VAE 150 times. (c) The standard deviation of the VAE output fields, calculated from the same set of fields as in (b).}
    \label{fig:VAE_of_truth}
\end{figure}

A notable characteristic of a VAE is that it produces different outputs when presented with the same input multiple times. In Figure~\ref{fig:VAE_of_truth}, we illustrate the mean and standard deviation of an ensemble of VAE outputs, reconstructed from the input $\teightfifty$ field for April 15, 2019. Similar to many machine learning approaches \citep[e.g.][]{Weyn2020a}, the VAE tends to smooth the reconstructed input field and does not capture all its features, particularly those at small spatial scales. These issues could be alleviated by extending the latent space vector or by fine-tuning the encoder-decoder architecture. 

Compared to the range of temperature anomalies for a given day of the year, the standard deviations of the output fields from the VAE were relatively small. To amplify them, a VAE should instead be trained using a smaller deterministic multiplier~$\deterministicmultiplier$, which would come at the expense of the reconstruction skill. Therefore, to achieve a more realistic spread of the output temperature field based on the climatological variability, we chose to replace $\bm{\sigma}_\phi$ from the VAE with the $\bm{\sigma}_b$ from the background-error covariance matrix (described in Section~\ref{sec:B modelling}) in the continuation of this study.

\subsubsection{VAE mappings}\label{section:mappings}

\begin{figure}[h!]
    \centering
    \includegraphics[width=\textwidth]{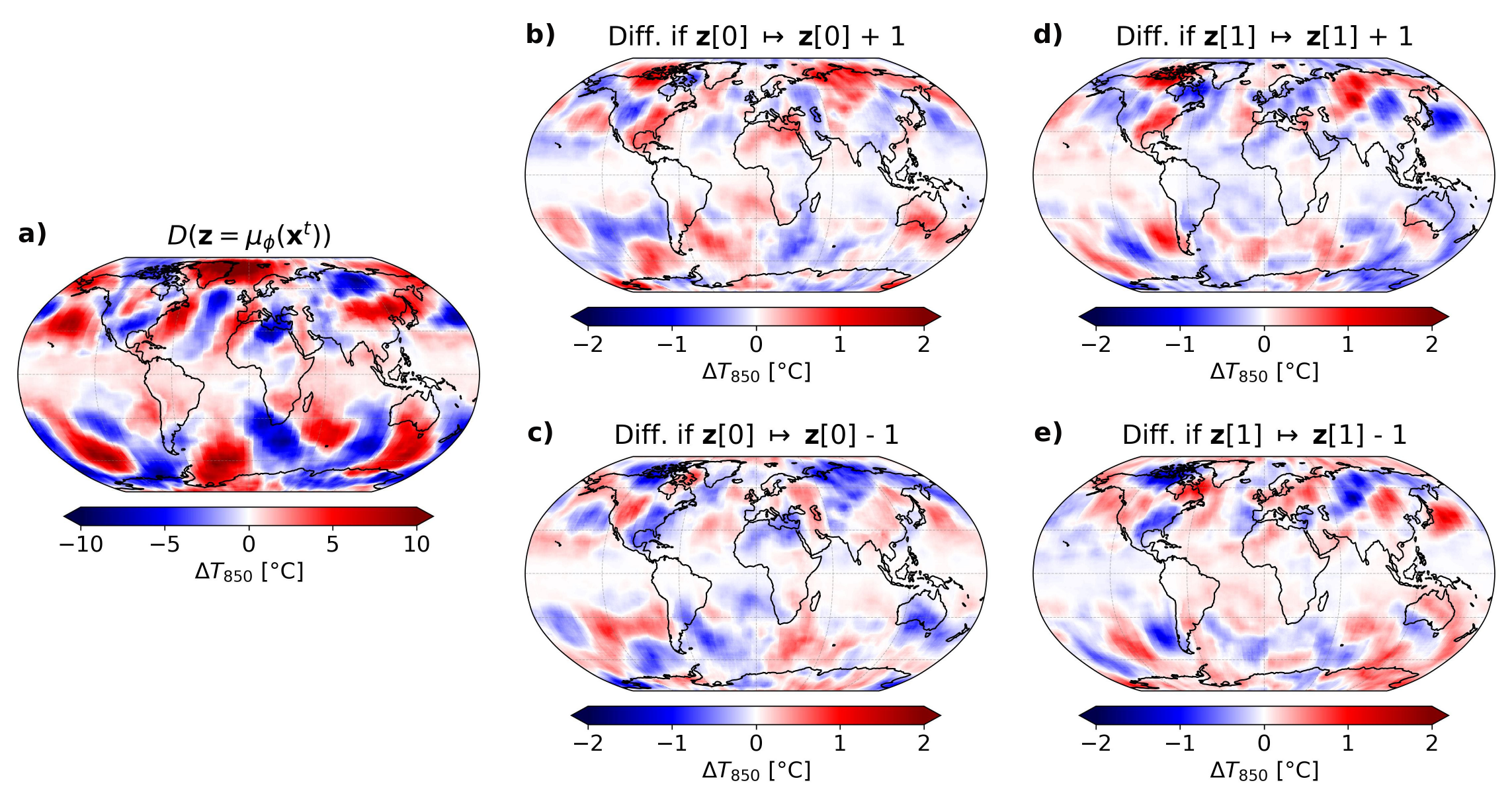}
    \caption{(a) Temperature at 850 hPa ($\mathbf{x}'$), reconstructed by the VAE from the ground truth ($\mathbf{x}^t$) for April 15, 2019, as $\mathbf{x}'=D(\mu_\phi(\mathbf{x}^t))$. (b) Difference to (a) if the first element of the latent vector is increased by~1. (c) Same as (b) but with the first element decreased by~1. (d,e) Same as (b,c), but with the second element of the latent vector being increased or decreased by~1. Note that the colour scales in (a) and (b-e) are different.}
    \label{fig:mappings}
\end{figure}

The decoder determines the mapping between the latent space and the grid point space. These mappings can be inspected by altering the original value of only one (say $j$-th) element of latent vector, $\mathbf{z}[j]$, to some different value, for example $\mathbf{z}[j]\mapsto \mathbf{z}[j]+1$. The original latent vector and the modified latent vector are then mapped to the grid point space using a decoder, and the difference in resulting grid point states describes the shape of the feature, represented by the $j$-th element of latent vector. Figure~\ref{fig:mappings} showcases two mappings to the grid-point space that correspond to the first two elements of the latent vector. These global patterns are rather complex and encapsulate both large-scale and small-scale features. Note that the mappings $\mathbf{z}[j]\mapsto \mathbf{z}[j]+1$ and $\mathbf{z}[j]\mapsto \mathbf{z}[j]-1$ are not exactly opposite in the grid point space due to nonlinearity of the decoder. 

The mappings vary daily due to (1) the seasonal variations of the climatological standard deviation, which is used in the destandardisation of the decoded field and due to (2) being state-vector dependent, a consequence of the nonlinearity of the decoder. Therefore, an equal perturbation of some latent vector element yields different responses for different values of the element. A more thorough description of the mapping is provided in Appendix~A.

\subsection{Data assimilation methodology}
\subsubsection{3D-Var-like data assimilation in the latent space}
\label{sec:da_latent_space_methodology}

3D-variational assimilation (3D-Var) seeks the state $\mathbf{x}$ of the atmosphere, which optimally combines the prior knowledge of the state denoted as \textit{background} ($\mathbf{x}_b$) and the \textit{observations} ($\mathbf{y}$) by minimising the cost function $\mathcal{J}(\mathbf{x})$, which measures the distance of the state $\mathbf{x}$ to the background ($\mathcal{J}_b$ term) and the observations 
($\mathcal{J}_o$ term) \citep{Lorenc1986, Kalnay2003}:
\begin{equation}
\begin{aligned}
\label{eq:3D-Var cost function}
    \mathcal{J}(\mathbf{x}) &= \mathcal{J}_b + \mathcal{J}_o =\\ &= \frac{1}{2} (\mathbf{x} - \mathbf{x}_b)^\mathrm{T} \mathbf{B}^{-1}(\mathbf{x} - \mathbf{x}_b) + \frac{1}{2}\left\{\mathbf{y} - H(\mathbf{x})\right\}^\mathrm{T}\mathbf{R}^{-1}\left\{\mathbf{y} - H(\mathbf{x})\right\} \, .
\end{aligned}
\end{equation}
$\mathbf{B}$ is the background-error covariance matrix, $\mathbf{R}$ is the observation-error covariance matrix, and $H$ is the observation operator, which produces an equivalent of the atmospheric state $\mathbf{x}$ in the observation space. The state which minimises the cost function (\ref{eq:3D-Var cost function}) is denoted an \textit{analysis}, $\mathbf{x}_a = \mathrm{arg}\,\min_{\mathbf{x}} \mathcal{J}(\mathbf{x})$.  

The 3D-Var cost function~(\ref{eq:3D-Var cost function}) is analytically derived by assuming that both the background and the observation errors are independent and Gaussian. We can assume that none of these assumptions is violated if the background state is defined in the latent space instead of the grid point space. Thus, we have defined cost function $\mathcal{J}_z$ in the reduced dimension latent space that measures the distance of the latent state $\latentvector$ to the background latent state $\latentvector_b$ (term $\mathcal{J}_{bz}$) and the distance of the observations $\mathbf{y}$ to the $\latentvector$, transformed into the observation space by $H (D(\mathbf{z}))$ (term $\mathcal{J}_{oz}$):
\begin{equation}\label{eq:3D-Var latent space}
\begin{aligned}
    \mathcal{J}_z(\latentvector) &= \mathcal{J}_{bz} + \mathcal{J}_{oz} = \\
    &= \frac{1}{2}(\latentvector - \latentvector_b)^\mathrm{T} \mathbf{B}_z^{-1}(\latentvector - \latentvector_b) + \frac{1}{2} \left\{\mathbf{y} - H(D(\latentvector))\right\}^\mathrm{T}\mathbf{R}^{-1}\left\{\mathbf{y} - H (D(\latentvector))\right\} \, .
\end{aligned}
\end{equation}
$\mathbf{B}_z$ is the background-error covariance matrix  in the latent space (described in Section~\ref{sec:B modelling}), $D$ stands for the decoder, and the observation operator $H$ interpolates the decoded field to the observation locations. The optimal latent space analysis $\mathbf{z}_a$ is obtained by minimising the cost function, such that $\mathbf{z}_a = \mathrm{arg}\,\min_{\mathbf{z}} \mathcal{J}_z(\mathbf{z})$. The minimisation algorithm, described in the Section~\ref{sec:minimisation}, computes the gradient of the cost function (\ref{eq:3D-Var latent space})
\begin{equation}\label{eq:cost_function_gradient}
    \nabla_z\mathcal{J}_z = \mathbf{B}_z^{-1} (\latentvector - \latentvector_b) + \mathbf{G}\,\mathbf{R}^{-1}\{\mathbf{y} - H(D(\latentvector))\} \, ,
\end{equation}
where $\mathbf{G}=\left(\partial H/\partial D \right) \partial D/ \partial\mathbf{z}$. The differentiation of $\mathcal{J}(\mathbf{z})$ is done automatically.

In operational NWP variational data assimilation, the main challenge for minimisation is the large dimension ($\sim$$10^9$) of the model state vector, which would result in $\mathbf{B}$-matrix with $\sim$$10^{18}$ elements. The minimisation of the cost function $\mathcal{J}$ would require computing its gradient $\nabla_\mathbf{x} \mathcal{J}$, and evaluating the inverse of $\mathbf{B}$, which would make the minimisation problem unfeasible. The issue is mostly tackled by 1) simplifying the background term $\mathcal{J}_b$ in (\ref{eq:3D-Var cost function}), and 2) performing minimisation in the space of reduced-dimension, i.e.~with reduced grid-point (or spectral) resolution. In our case, we have decreased the expense of minimisation enormously by transforming the minimisation problem to the low-dimension latent space. For example, the state vector $\mathbf{x}$ has $720\times1440=1\,036\,800$ elements, which results in more than $10^{12}$ elements in~$\mathbf{B}$. On the other hand, the latent vector $\mathbf{z}$ has 100 elements, with full $\mathbf{B}_z$-matrix containing only $10^4$ elements. 

To simplify the computation of the background term, operational DA employs additional assumptions regarding the structure and properties of the $\mathbf{B}$-matrix. These include assumptions about 1) the spatial relations of background errors, for example spatial homogeneity of their correlation length-scale, and about 2) the physical relations (balances) of background errors \citep[for a review of the topic see][]{Bannister2008a,Bannister2021}. These relations can be described by simplified diagnostic equations of the atmospheric flow, such as the linearised nonlinear balance equation, quasi-geostrophic omega equation, and thermodynamic balances \citep{ecmwf_da_2023}. Their main role is to provide balanced initial conditions, so that the information from the assimilated observations does not rapidly disappear and propagate away as gravity waves. These assumptions are used to make a change of variable, commonly referred to as the control variable transform (CVT), which converts the assimilation increment $\delta\mathbf{x}=\mathbf{x}-\mathbf{x}_b$ into a \textit{control}-$\chi$ \textit{space}, such that $\delta\mathbf{x}=\mathbf{B}^{1/2}\chi$. In the $\chi$-space, the background-error covariance matrix becomes an identity matrix, so the DA problem becomes feasible. The transformation effectively decreases the condition number of the Hessian matrix $\partial^2 \mathcal{J}/\partial \mathbf{x}^2$, and consequently improves the convergence rate of minimisation. 

In the next section, we will use the VAE discovered transformation between the model space and latent space to model the $\mathbf{B}_z$-matrix.

\subsubsection{Background-error covariance modelling}
\label{sec:B modelling}

\begin{figure}[h!]
    \centering
    \includegraphics[width=\textwidth, clip, trim={0cm 0.4cm 0cm 0cm}]{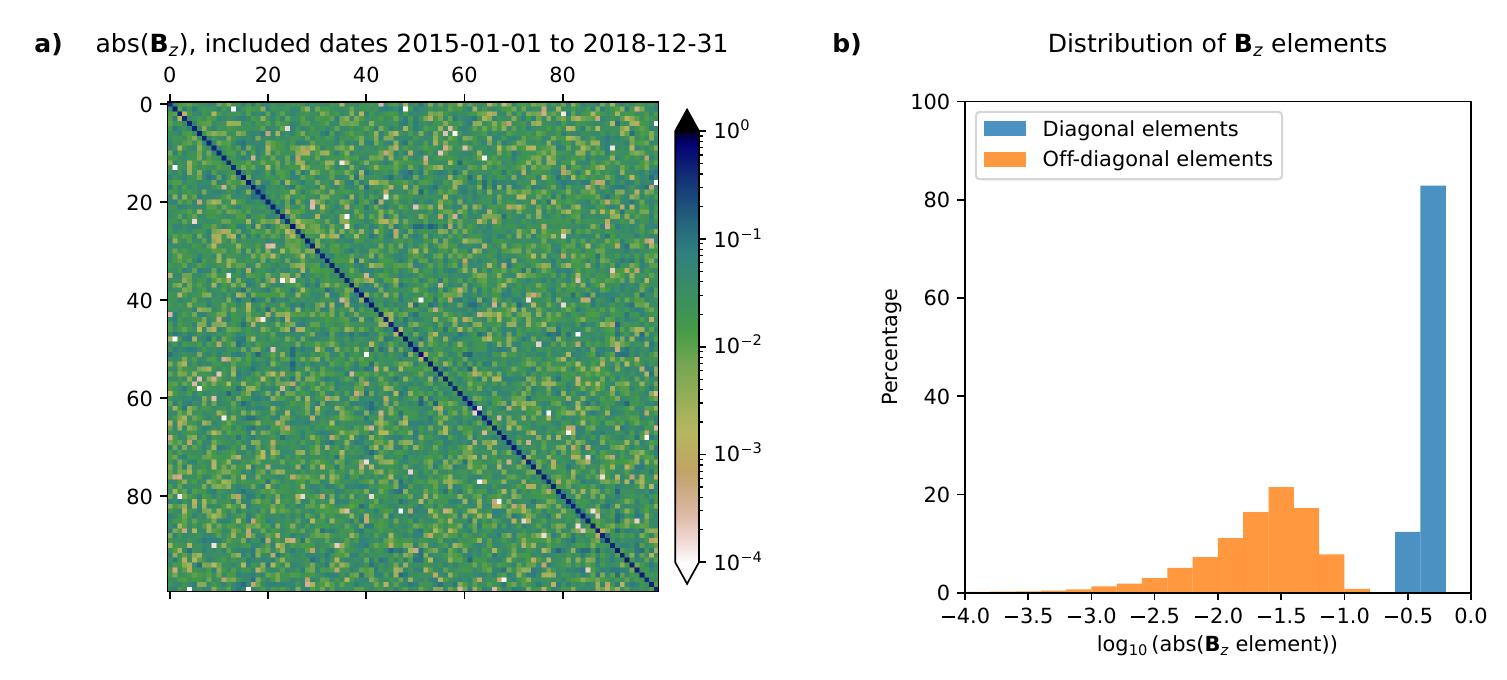}
    \caption{(a) Background-error covariance matrix $\mathbf{B}_z$ represented in the latent space for the validation set. The absolute values of $\mathbf{B}_z$-matrix elements are shown. The diagonal elements represent the background-error variances, and off-diagonal elements represent the error covariances of different elements of the background latent vector. (b) Distribution of the values of the diagonal and off-diagonal elements in (a). The bin width in logarithmic scale is $0.2$.}
    \label{fig:B matrix classic}
\end{figure}

The forward model providing the background ($\mathbf{z}_b$) for the assimilation is based on persistence in the latent space (see description in Section~\ref{sec:osse}, Equation~(\ref{eq:persistence})). Therefore, the covariance matrix for background errors in the latent space, $\mathbf{B}_z$, was computed by taking climatological covariance statistics of the differences between pairs of unperturbed ($\mathbf{z}=\bm{\mu}_{\phi}$) encoded ground truth states on two consecutive days ($\latentvector_t^{d}$ and $\latentvector_t^{d-1}$), while taking into account zero bias of the persistence forecast in the latent space: 
\begin{equation}
\label{eq:B classic}
\begin{aligned}
    \mathbf{B}_z &= \left<\left(\latentvector_b - \latentvector_t\right)\,\left(\latentvector_b - \latentvector_t\right)^\mathrm{T}\right> \\
    &=\left< \left(\latentvector_t^{d-1} - \latentvector_t^{d}\right)\, \left(\latentvector_t^{d-1} - \latentvector_t^{d}\right)^T \right>  \, .
\end{aligned}
\end{equation}
Latent vector $\latentvector_t$ is the unperturbed encoded ground truth, $\latentvector_b$ stands for the encoded background, and $\left<\cdot\right>$ denotes averaging over multiple ($\latentvector_t^{d}$, $\latentvector_t^{d-1}$) pairs. The diagnosed $\mathbf{B}_z$ for the validation set is shown in Figure~\ref{fig:B matrix classic}. It is quasi-diagonal, with diagonal elements mostly more than an order of magnitude larger than the off-diagonal elements (Figure~\ref{fig:B matrix classic}b). If we thereby assume a diagonal $\mathbf{B}_z$, the background term in (\ref{eq:3D-Var latent space}) would be already significantly simplified. Therefore, performing a CVT would only require dividing the latent state $\mathbf{z}$ element-wise by the square root of the background-error variances in the latent space, i.e. $\mathrm{diag}\left(\mathbf{B}_z^{1/2}\right)$, while the VAE has already captured the spatial relations in the $\teightfifty$ field and its associated error field. Consequently, our latent space closely resembles the control space. The primary distinction lies in the fact that, in our case, the transformation from the model space to the latent space is learned from historical data, whereas the traditional transformation to the control space is based on assumptions and the manner in which $\mathbf{B}^{1/2}$ is modelled.

Computing an inverse of a full $\latentB$ matrix with $100\times100$ elements is cheap in our case, but we opted for an approach required in the case of a large latent space (as in operational data assimilation), and only retained the large diagonal elements of $\latentB$, so that $\latentB$ is easy to invert. The term  $\mathbf{B}_z^{-1} \left(\mathbf{z}-\mathbf{z}_b\right)$ in (\ref{eq:cost_function_gradient}) is then replaced by a simple element-wise division of $\left(\mathbf{z}-\mathbf{z}_b\right)$ with the vector of background-error variances in the latent space $\bm{\sigma}_b^2=\mathrm{diag}\left(\mathbf{B}_z\right)$:
\begin{equation*}
    \mathbf{B}_z^{-1} \left(\mathbf{z}-\mathbf{z}_b\right)\approx (\mathbf{z}-\mathbf{z}_b)\oslash \bm{\sigma}_b^2 \, .
\end{equation*}
 This further accelerates the minimisation. In the experiments we conducted, the $\mathbf{B}_z$-matrix was assumed diagonal, except if stated otherwise. The validity of using only diagonal elements is further discussed in Section~\ref{sec:midlatitudes}.

\begin{figure}[h!]
    \centering
    \includegraphics[width=\textwidth]{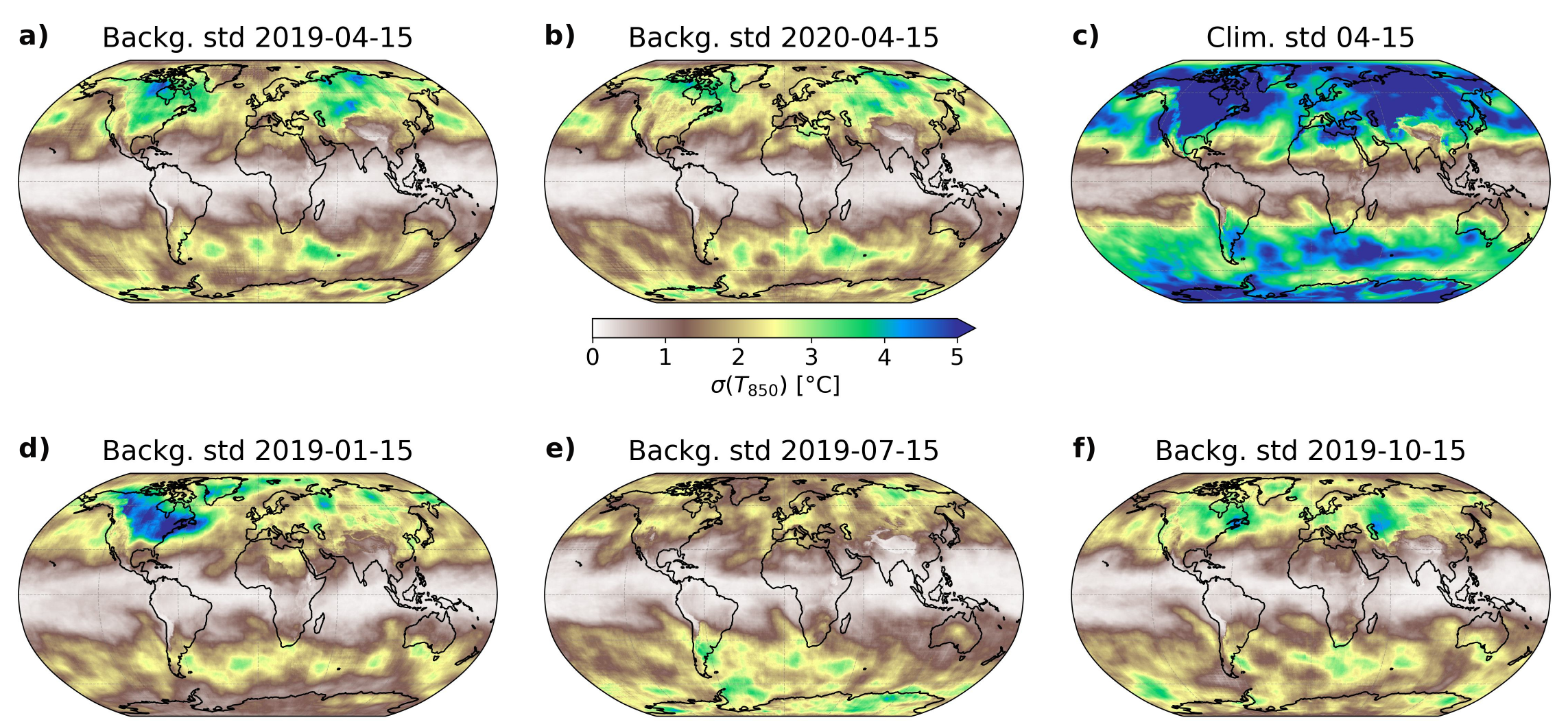}
    \caption{Background-error standard deviation for different dates and the climatological standard deviation for April~15. The colourbar in the centre (unit $\SI{}{\degreeCelsius}$) applies to all panels. The background-error standard deviations were computed from 150 decoded ensemble members.}
    \label{fig:background std}
\end{figure}

Although~$\mathbf{B}_z$ was constant in the latent space, the background-error standard deviation in the grid point space varied daily, as shown in Figure~\ref{fig:background std}. This variability can be attributed to 1) variations in the climatological standard deviation of each day of the year, which is used for the destandardisation of the decoded field (recall description in Section~\ref{sec:data}), and 2) differences in the mean state of the background latent vector $\mathbf{z}_b$ and the nonlinearity of the decoder. Because of 1), the background-error standard deviations differ between seasons (Figures~\ref{fig:background std}a,d-f). As expected, they are greater in the winter hemisphere and over the continents. Because of 2), the background-error standard deviations differ in a flow-dependent way, i.e.~based on the background latent state. For example, compare the background-error standard deviations over North America for the same date but different year, illustrated in Figures~\ref{fig:background std}a,b. In Appendix~A, we prove, that the differences in the background error standard deviations between Figures~\ref{fig:background std}a and \ref{fig:background std}b are a consequence of different background latent state and the nonlinearity of the VAE decoder, and not a consequence of the use of a finite ensemble.

\subsubsection{Minimisation algorithm}\label{sec:minimisation}
To minimise the cost function~\eqref{eq:3D-Var latent space}, we used the Adam optimizer \citep{kingma2017adam}. For single observation experiments, the initial learning rate was set to $0.01$, while for other experiments the learning rate was set to $0.1$. The other properties of the Adam optimizer were set to their default settings as implemented in TensorFlow. Note that the weights of the decoder and encoder were not allowed to change during the minimisation of the DA cost function~(\ref{eq:3D-Var latent space}).

During the minimisation process, if the value of the cost function did not improve within the last three steps, the learning rate was reduced by a factor of $2$ but not allowed to drop below $10^{-4}$. The minimisation was terminated if the criterion
\begin{equation}\label{eq:stopping_criterion}
    1 - \frac{\mathcal{J}_z\left(\latentvector^{(i)}\right)}{\mathcal{J}_z\left(\latentvector^{(i-1)}\right)} < \varepsilon
\end{equation}
was fulfilled for 10 consecutive steps. Here $\mathbf{z}^{(i)}$ represents the latent vector at the $i$-th iteration, and the threshold $\varepsilon$ was set to $0.01$. Alternatively, the algorithm would automatically terminate if it reached 100 computational steps. The $\latentvector$ corresponding to the lowest $\mathcal{J}_z$ value was always chosen as the analysis. Typically, the minimisation converged in 10 to 30 iterations, while a single minimisation process lasted a couple of seconds on a single CPU.

\subsubsection{Setup of observing system simulation experiments}\label{sec:osse}

\begin{figure}[h!]
    \centering
    \includegraphics[width=0.8\textwidth]{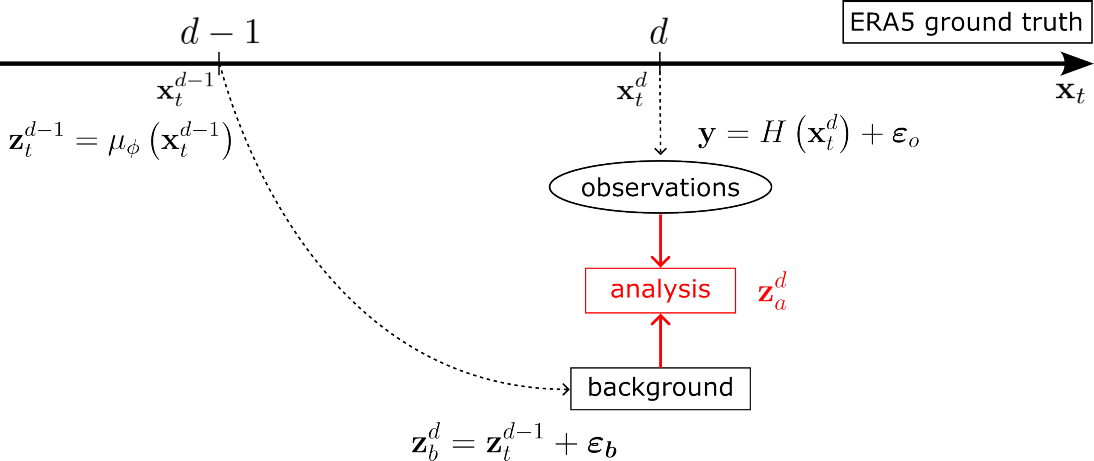}
    \caption{Setup of the observing system simulation experiments with neural network data assimilation system. Both observations and background are generated from ERA5 ground truth $\mathbf{x}_t$. The observations $\mathbf{y}$ on day $d$ are fused with the latent-space background $\mathbf{z}_b^d$, that is a perturbed 1-day persistence-based forecast issued on day $d-1$, to form the latent space analysis $\mathbf{z}_a$.}
    \label{fig:osse}
\end{figure}

We performed a series of observing-system simulation experiments (OSSEs). For simplicity, we generated both the background and the observations from ERA5 ground truth, as depicted in Figure~\ref{fig:osse}, and computed the analysis following the minimisation of the cost function (\ref{eq:3D-Var latent space}). 

The observations on day $d$ were simulated by adding random Gaussian perturbations with zero mean and standard deviation $\bm{\sigma}_o$ (i.e.~$\bm{\varepsilon}_o\sim\mathcal{N}(\bm{0},\bm{\sigma}_o)$) to the ground truth values $\mathbf{x}_t^d$, bilinearly interpolated with $H$ to the location of the observations, $H\left(\mathbf{x}_t^d\right)$, which yields the observation vector:
\begin{equation}\label{eq:eda_obs}
    \mathbf{y} = H\left(\mathbf{x}_t^d\right) + \bm{\varepsilon}_o \, .
\end{equation}
The observations were assumed uncorrelated and their error standard deviation was set to 1 K in all experiments.

For simplicity, we chose the background latent state to remain constant in time, so the $\mathbf{z}_b^d$ on day $d$ was equal to the encoded mean ground truth of the previous day, $d-1$, i.e.,
\begin{equation}\label{eq:persistence}
    \mathbf{z}_b^d = \mathbf{z}_t^{d-1} = \mu_\phi\left(\mathbf{x}_t^{d-1}\right) \, .
\end{equation}
This type of model is commonly referred to as a persistence model.  We perturbed the $\latentvector_b^d$ by adding Gaussian noise $\bm{\varepsilon}_b\sim\mathcal{N}\left(\bm{0},\bm{\sigma}_b\right)$ with standard deviation $\bm{\sigma}_b=\mathrm{diag}\left(\mathbf{B}_z^{1/2}\right)$:
\begin{equation}\label{eq:eda_bg}
    \mathbf{z}_b^d = \mu_\phi\left(\mathbf{x}_t^{d-1}\right) + \bm{\varepsilon}_b \, .
\end{equation}

Due to the efficiency of the computations, we conducted all experiments using an ensemble of 150 variational data assimilations, i.e.~by repeating the data assimilation procedure (Equations~(\ref{eq:3D-Var latent space}) and (\ref{eq:cost_function_gradient})) multiple times. We compute each analysis based on the procedure described above from a pair of perturbed $\mathbf{y}$ (based on the $\mathbf{R}$-matrix, Equation~(\ref{eq:eda_obs})) and perturbed $\mathbf{z}_b$ (based on the climatological background-error covariance matrix $\mathbf{B}_z$, Equation~(\ref{eq:eda_bg})). This way, we obtain 150 analyses in the latent space, which are transformed to the grid point space, where we compute their statistics, e.g. the analysis-error standard deviation $\sigma_a$.

Our approach somewhat resembles the traditional ensemble of data assimilations (EDA) \citep{Isaksen2010,Bonavita2012}, however, our persistence-based forward model does not produce the so-called error of the day, i.e.~the \textit{fully} flow-dependent background error variances, corresponding to the current atmospheric conditions.

\section{Results}\label{section:results}

\subsection{Single observation experiments}
\label{sec:singobs}
The background-error model differs significantly between the tropics and the midlatitudes due to variations in atmospheric dynamics, primarily attributed to the latitude-varying Coriolis force. In the midlatitudes at synoptic scales, the prevailing thermal wind balance is continuously restored. Conversely, in equatorial areas, the main balance is between the vertical motions and the condensational heating. The latter leads to the excitation of equatorial waves, which propagate within the equatorial wave channel \citep{Matsuno1966}, and effectively couple the remote tropical areas. Previous studies have explored different approaches to represent background-error covariances in these regions. For instance, \citet{Zagar2004a,Zagar2005} utilised equatorial waves as basis functions in a spectral representation of tropical background-error covariances, and \citet{Kornich2008} combined the midlatitude and equatorial balances using Hough modes at certain characteristic depth of atmospheric features by extending the control vector for minimisation. Notably, all these attempts were carried out within a shallow-water model framework. Subsequently, \citet{Zagar2013} used normal modes of linearised atmospheric motions \citep{Kasahara1981} to diagnose short-range forecast covariances from the ECMWF ensemble forecasts. However, in the operational data assimilation at ECMWF, there is no dedicated $\mathbf{B}$-model for the tropics, and the assimilation of dynamic variables is univariate. In the midlatitudes, on the other hand, the balance is defined by the linearised version of nonlinear balance equation and quasi-geostrophic $\omega$ equation \citep{ecmwf_da_2023}.

In this study, we present a unified model for the background-error covariances in the tropics and midlatitudes and show that the neural-network-derived mappings effectively capture the spatial autocovariances of the $\teightfifty$ background errors. The background-error model is demonstrated in the following sections through several single observation experiments. Each of our single observation data assimilation experiments is an ensemble of 150 data assimilations, which allows us to estimate the analysis-error standard deviation and compare it to the background-error standard deviation. For simplicity, we will denote ensemble-mean background as \textit{background}, ensemble-mean analysis as \textit{analysis}, ensemble-mean analysis increment (i.e.~ensemble-mean analysis minus ensemble-mean background) as \textit{analysis increment} and mean VAE-reconstructed truth as \textit{reconstructed truth} throughout Section~\ref{section:results}.

\subsubsection{Observation in midlatitudes}
\label{sec:midlatitudes}
\begin{figure}[h!]
    \centering
    \includegraphics[width=\textwidth]{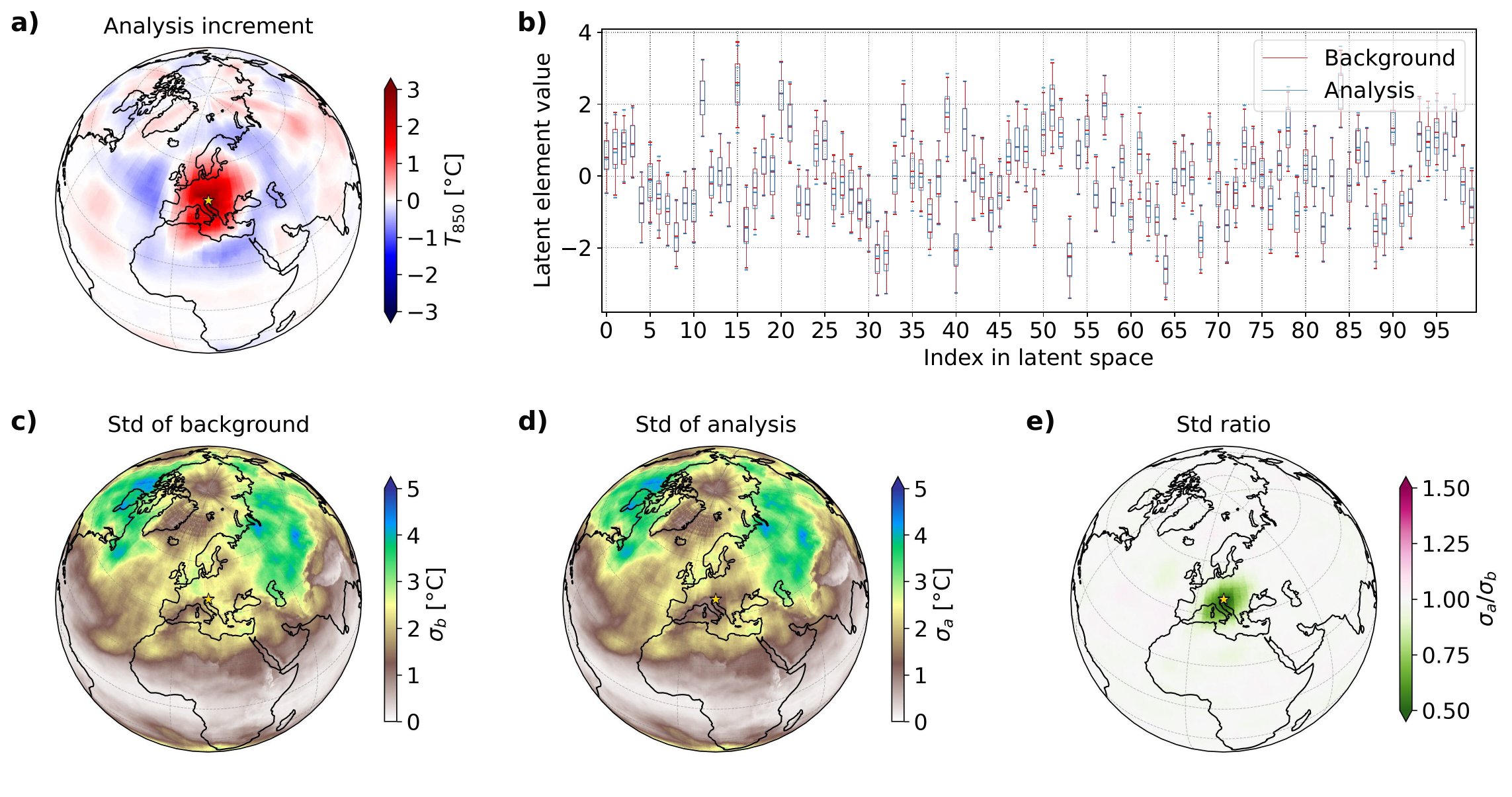}  
    \caption{Single observation experiment above Ljubljana with a 3\,K departure and $\sigma_o$ of 1\,K, utilising the background temperature field for April 15, 2019. (a) Mean analysis increment (analysis mean minus background mean). (b) Latent vector distribution for background (red) and analysis (blue) with boxes spanning 25th and 75th percentiles and whiskers from the 5th to 95th percentile. Outliers are not shown. (c) Background-error standard deviation, $\sigma_b$. (d) Analysis-error standard deviation, $\sigma_a$. (e) Ratio $\sigma_a/\sigma_b$. The observation location is marked with a golden star in panels (a,c-e).}
    \label{fig:singobs_Ljubljana}
\end{figure}

The first experiment demonstrates the impact of a single temperature observation with 3\,K observation departure $\mathbf{d}=\mathbf{y}-H(D(\mathbf{z}_b^d))$ and a standard deviation $\sigma_o$ of 1\,K. The observation is located above Ljubljana, Slovenia, at $46.1\,^\circ\mathrm{N}$ and $14.5\,^\circ\mathrm{E}$. Figure~\ref{fig:singobs_Ljubljana}a shows that the ensemble-mean analysis increment following an ensemble of data assimilations has its expected structure with respect to the typical atmospheric flow. Firstly, the increment peaks at the observation point, while it is also large over the continental part (Alps, Balkans and the central Europe), where the background-error standard deviation is relatively larger than over the Mediterranean (Figure~\ref{fig:singobs_Ljubljana}c). This is expected, as the climatological temperature variance over the Mediterranean is lesser than over the continent due to the damping effect of sea surface temperature on $\teightfifty$ variability. Secondly, the shape of the positive increment is elongated in the south-west to north-east direction, which complies with the mean south-westerly winds in the region. Thirdly, the area of positive increment is surrounded by a shallower negative increment. Again, this is an expected result for the applied climatological background-error covariance model \citep{Fisher2003}, as the negative temperature perturbation can be associated with a spatial translation of a synoptic Rossby wave in the background with respect to reality. The amplitude of the increment becomes much smaller further away from the observation location. 

Panel~\ref{fig:singobs_Ljubljana}d shows the resulting analysis-error standard deviation $\sigma_a$ after performing an ensemble of data assimilations. A significant reduction of $\sigma_a$ with respect to $\sigma_b$ is only observed in the proximity of the observation location with a few hundred-kilometre elongation towards the southwest in compliance with the typical south-westerlies (panel \ref{fig:singobs_Ljubljana}e, also compare panels \ref{fig:singobs_Ljubljana}c and \ref{fig:singobs_Ljubljana}d). Consistently with small and localised changes in the grid point space, the changes in the latent state are also barely visible (panel \ref{fig:singobs_Ljubljana}b). For example, the average standard deviation of the elements of the background latent state was 0.665, while the average standard deviation of the elements of the analysis latent state was 0.662. The analysis standard deviation is smaller than the background standard deviation for 69 out of 100 latent vector elements. 

\begin{figure}[h!]
    \centering
    \includegraphics[width=\textwidth]{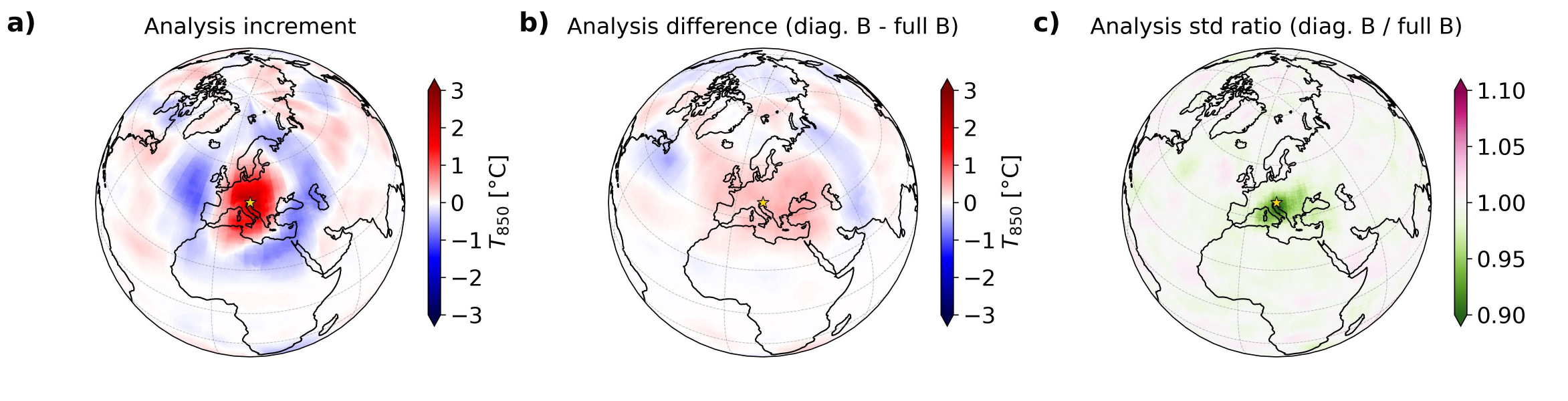}
    \caption{Assimilation of a single temperature observation above Ljubljana, as in Figure~\ref{fig:singobs_Ljubljana}, but with the full $\latentB$ used for the assimilation. (a) The analysis increment. (b) The difference between the analyses here and in Figure~\ref{fig:singobs_Ljubljana}. (c) The ratio between the standard deviations of the analyses from Figure~\ref{fig:singobs_Ljubljana} and the analysis for this experiment.}
\label{fig:singobs_Ljubljana_full_B}
\end{figure}

 We justify the use of diagonal $\latentB$ for data assimilation by comparing the analysis increment with diagonal $\latentB$ (Figure~\ref{fig:singobs_Ljubljana}a) to the analysis increment with full $\latentB$, shown in Figure~\ref{fig:singobs_Ljubljana_full_B}a. The observation has a slightly larger impact in the case of the diagonal $\latentB$ (Figure~\ref{fig:singobs_Ljubljana}b).
 Correspondingly, the analysis-error standard deviation in the case of the diagonal $\latentB$ is also somewhat smaller in the proximity of the observation (the ratio in Figure~\ref{fig:singobs_Ljubljana}c is everywhere between 0.88 and 1.03).
Nevertheless, the differences in the analyses and their standard deviations are significantly smaller than the impacts of the observation on the analyses and their standard deviations. 
 Thus we can conclude that using only the diagonal elements of $\latentB$ instead of full $\latentB$ does not importantly harm the quality of the data assimilation.

\begin{figure}[h!]
    \centering
    \includegraphics[width=\textwidth]{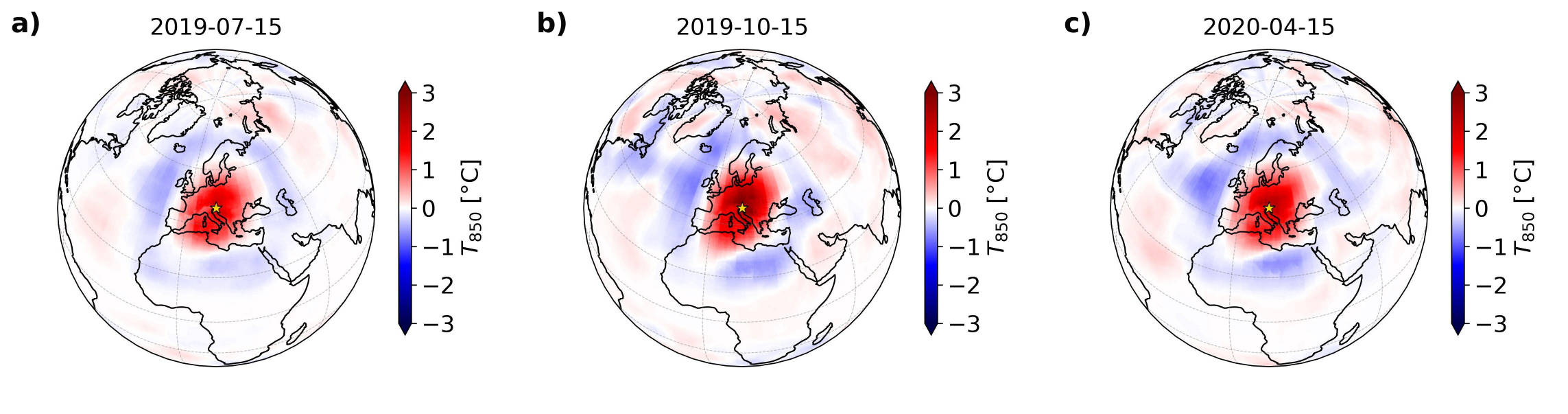}
    \caption{Analysis increments for single observation experiments for an observation above Ljubljana (as in Figure~\ref{fig:singobs_Ljubljana}) on different dates: a) July~15, 2019; b) October~15, 2019; c) April 15, 2020.}
\label{fig:singobs_Ljubljana_other_dates}
\end{figure}

Another important property of data assimilation is that the analysis increment does not only depend on the observation departure but also on the structure and magnitude of the the background-error covariances, that are in our case mildly-dependent on $\mathbf{z}_b$, i.e. flow-dependent, as described in Section~\ref{sec:osse}. Figure~\ref{fig:singobs_Ljubljana_other_dates} shows the analysis increments for the single observation experiments with the same temperature departure at the same location as in Figure~\ref{fig:singobs_Ljubljana}, however for different dates, and so different backgrounds, and background-error covariances. The differences between the increments are small but clearly visible. 

\begin{figure}[h!]
    \centering
    \includegraphics[width=\textwidth]{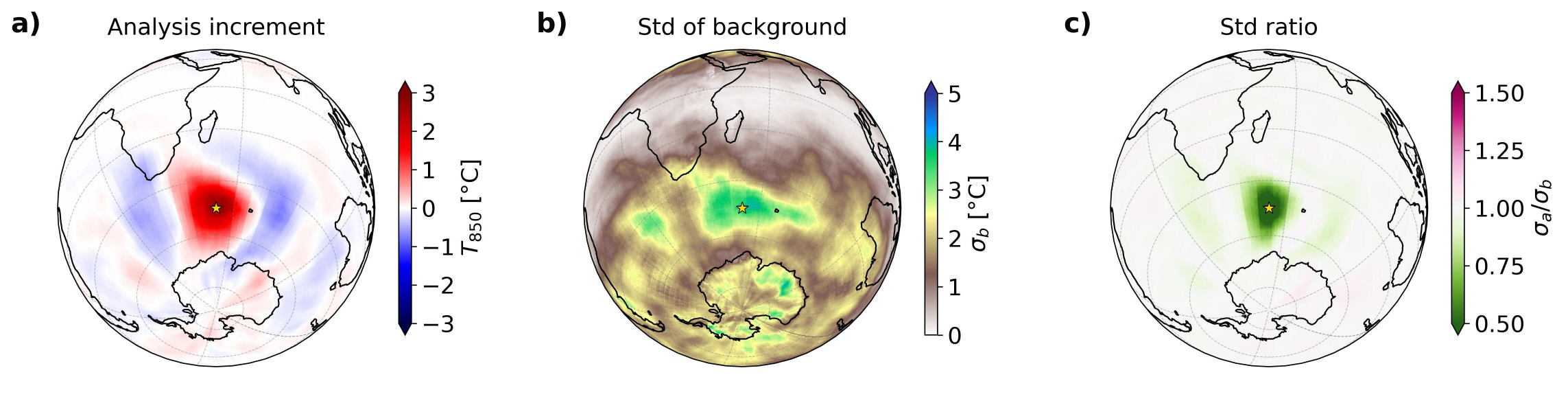}
    \caption{Single observation experiment for a temperature observation above southwestern Indian Ocean with a 3\,K departure and $\sigma_o$ of 1\,K, with the background temperature field for April 15, 2019. (a) The mean analysis increment. (b) Standard deviation of the background, $\sigma_b$. (c) The ratio between the standard deviations of the analysis and of the background, $\sigma_a/\sigma_b$. The golden star marks the observation location.}
    \label{fig:singobs_SW_Indian_Ocean}
\end{figure}

Therefore, the change in the observation location should not only result in the spatial translation of the increment, but should also be sensitive to the $\mathbf{z}_b$ and associated patterns of $\sigma_b$.
Figure~\ref{fig:singobs_SW_Indian_Ocean} shows an example of assimilation of a single observation with a 3\,K observation departure and $\sigma_o$ of 1\,K. The observation is located near the centre of the area with high $\sigma_b$ in the southwestern Indian Ocean at $50\,^\circ\mathrm{S}, 50\,^\circ\mathrm{E}$ (Figure~\ref{fig:singobs_SW_Indian_Ocean}b). Due to a large background-error standard deviation, the information coming from the observation is weighed more than the background information, resulting in an analysis increment of large magnitude, and strong reduction of $\sigma_a/\sigma_b$ (Figure~\ref{fig:singobs_SW_Indian_Ocean}c). The shape of the increment is anisotropic and extended towards the north-west, upstream of the temperature advection.  

\subsubsection{Observation in tropics}
As noted in Section~\ref{sec:singobs}, distinct atmospheric dynamics between the tropics and midlatitudes lead to contrasting background-error covariances. The correlation lengthscale is extended in the tropics, as the equatorial waves couple the remote areas, affecting analysis increments' shapes.

\begin{figure}[h!]
    \centering
    \includegraphics[width=\textwidth]{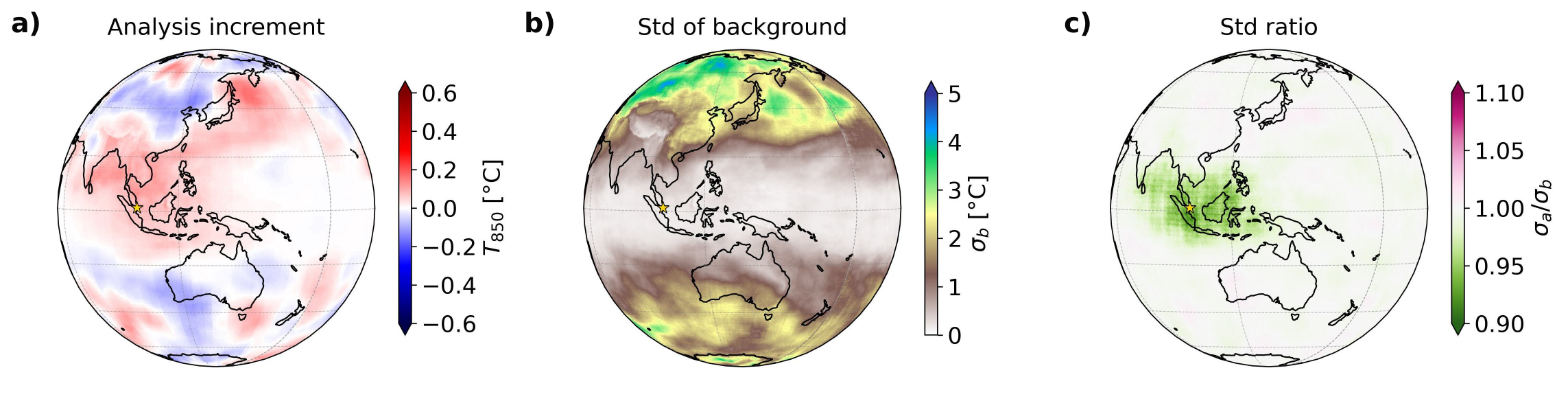}
    \caption{As Figure~\ref{fig:singobs_SW_Indian_Ocean}, but for an observation above Singapore. Note that the colour scales in (a,c) are different to those in previous experiments.}
    \label{fig:singobs_Singapore}
\end{figure}

The settings for the assimilation experiments with a single temperature observation were the same as for the midlatitudes. The observation, positioned above Singapore at $1.3\,^\circ\mathrm{N}, 103.9\,^\circ\mathrm{E}$ has a temperature departure of 3\,K and $\sigma_o$ of 1\,K. The resulting analysis increment is shown in Figure~\ref{fig:singobs_Singapore}a. We observe several notable properties. Firstly, the magnitude of the analysis increment at the observation location is small, as $\sigma_o\gg\sigma_b$, with $\sigma_b\approx0.19$\,K. A small analysis increment in the grid point space is thus expected. 
Secondly, there are significant temperature increments in the midlatitudes. These stem from spatially extensive background error correlations in the tropics, excessive $\sigma_b$ in the midlatitudes, and underestimated $\sigma_b$ in the tropics, as commonly observed in  climatology-based estimation of background-error covariances \citep{Bannister2008}. The midlatitudes are thus only weakly constrained by the background. 
This increment pattern shares similarities  with the temperature perturbation response over the Maritime continent \citep[e.g.][]{Trenberth1998,Kosovelj2019}, where condensational heating generates a divergent pattern exciting Rossby wave train along the great circle, particularly evident over eastern Asia towards Japan and North America. The effect is, as in our case, amplified in the Northern Hemisphere due to low-latitude subtropical jet in the vicinity of the Himalayas. This means there is a solid physical reasoning for such pattern. Outside the tropics, the analysis increment is an order of magnitude smaller than $\sigma_b$, and the assimilation only leads to a slight reduction of $\sigma_a/\sigma_b$ (Figure~\ref{fig:singobs_Singapore}b,c). Additionally, the analysis increment extends slightly eastwards from the observation location (also Figure~\ref{fig:singobs_Singapore}a,c), corresponding to the mean easterly winds in the tropical lower troposphere.

\begin{figure}[h!]
    \centering
    \includegraphics[width=\textwidth]{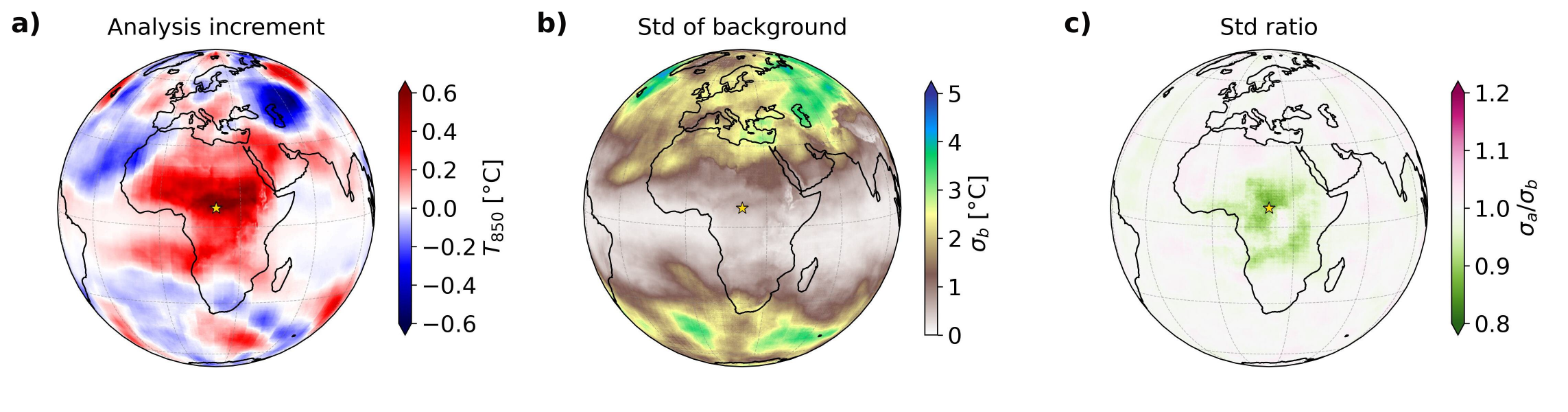}
    \caption{As Figure~\ref{fig:singobs_SW_Indian_Ocean}, but for an observation above central equatorial Africa. Note that the colour scales in (a,c) are different to those in previous experiments.}
    \label{fig:singobs_CAR}
\end{figure}

Figure~\ref{fig:singobs_CAR} provides another example featuring an observation in equatorial Africa at $7.0\,^\circ\mathrm{N}$, $21.0\,^\circ\mathrm{E}$, with $\sigma_b\approx\sigma_o$. The magnitude of the analysis increment at the observation location is larger than in the previous case. In the tropics and subtropics, the shape of the increment resembles the Equatorial Rossby wave pattern with symmetric positive increments to the north and south of the Equator, west of the observation location. The increment's magnitude in these regions could also be attributed to large $\sigma_b$ (Figure~\ref{fig:singobs_CAR}b). Again, the analysis increment includes the Rossby wave train, that can be observed over the Caspian Sea (Figure~\ref{fig:singobs_CAR}a).

\begin{figure}[h!]
    \centering
    \includegraphics[width=\textwidth]{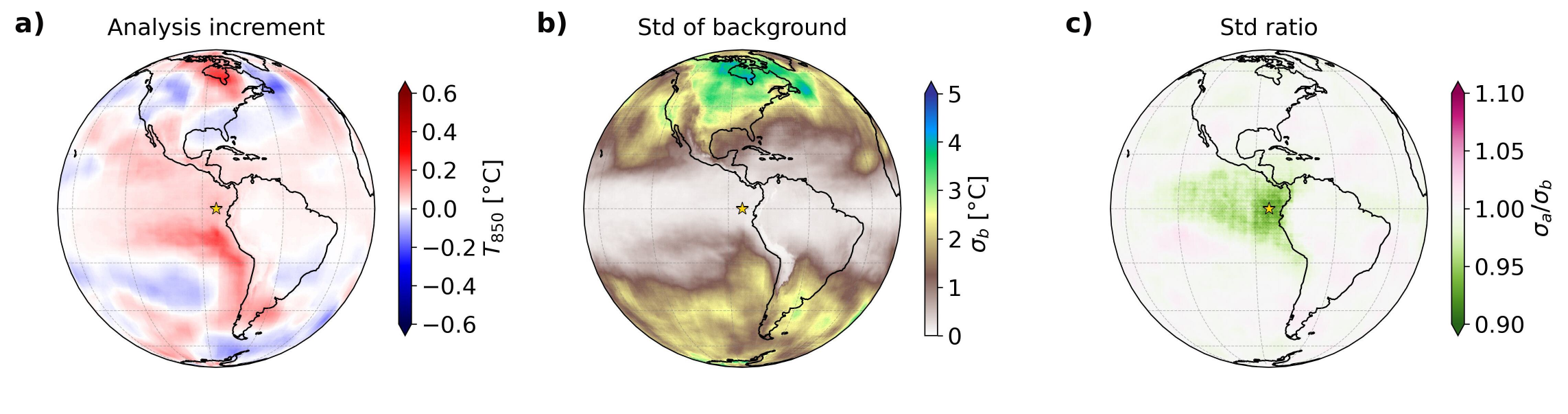}
    \caption{As Figure~\ref{fig:singobs_Ljubljana}, but for an observation above Eastern Pacific. Note that the colour scales in (a,c) are different to those in previous experiments.}
    \label{fig:singobs_E_Pacific}
\end{figure}

The final example in Figure~\ref{fig:singobs_E_Pacific} features an observation above the East Pacific at the Equator, $85.0\,^\circ\mathrm{W}$. As in Figure~\ref{fig:singobs_Singapore}, $\sigma_o\gg\sigma_b$, so the analysis increment is of small magnitude. The shape of the analysis increment exhibits an ENSO-like pattern with an amplified increment along the Peruvian coast. The reduction of $\sigma_a/\sigma_b$ extends far towards the central Pacific, corresponding to the strongly correlated lower branch of the Pacific Walker circulation. It also extends far into the subtropics, corresponding to the lower return branch of the Hadley circulation. However, the observation impact does not reach over the Andes into the Amazon. The extended correlation length scale over the tropical oceans and the effect of the orographic barrier on the analysis increment are further demonstrations of the capabilities of the NN-discovered transformations in the background-error covariance model.

Having performed a plethora of single observation experiments, we have also tested whether the results satisfy an important theoretical property of a single observation DA, that the analysis increment at the observation location $\delta\teightfifty^a$ and its standard deviation $\sigma_a$ are defined by:
\begin{equation}
    \label{eq:theoretical increment}
    \delta\teightfifty^a = \frac{\delta\teightfifty^o / \sigma_o^2}{1/\sigma_b^2 + 1/\sigma_o^2}
    \qquad\text{and}\qquad
    \sigma_a = \sqrt{\frac{1}{1/\sigma_b^2 + 1/\sigma_o^2}},
\end{equation}
where $\delta\teightfifty^o$ signifies the observation departure. The outcomes of this evaluation for the presented experiments are listed in Table~\ref{tab:singobs increments}. Generally, the experimental values closely align with the theoretical expectations. Minor discrepancies were likely stem from finite ensemble members and an imperfect minimisation algorithm.

\begin{table}[h!]
\caption{Comparison of the results from single observation experiments presented above with 150 ensemble members and preset $\delta\teightfifty^o=3\,\mathrm{K}$ and $\sigma_o=1\,\mathrm{K}$ with the theoretical expectations based on Equation~\eqref{eq:theoretical increment}. All values are in~K. The first two columns display deviations from preset values caused by finite ensemble sizes. Theoretically predicted values of $\delta\teightfifty^a$ and $\sigma_a$ are labelled as \textit{Theo.}, while experimental values carry the label \textit{Exp.} The values for $\sigma_b$, $\delta\teightfifty^a$, and $\sigma_a$ were obtained by bilinear interpolation from the grid to the observation location.}
\label{tab:singobs increments}
\centering
\begin{tabular}{lccccccc}
\toprule
Location & $\boldsymbol{\delta\teightfifty^o}$ & $\boldsymbol{\sigma_o}$ & $\boldsymbol{\sigma_b}$ & Theo.~$\boldsymbol{\delta\teightfifty^a}$ & Ex.~$\boldsymbol{\delta\teightfifty^a}$ & Theo.~$\boldsymbol{\sigma_a}$ & Ex.~$\boldsymbol{\sigma_a}$\\
\midrule
Ljubljana & 3.03 & 1.07 & 1.91 & 2.31 & 2.19 & 0.93 & 0.94\\
SW Indian Ocean & 3.14 & 0.95 & 3.86 & 2.96 & 2.95 & 0.92 & 0.95\\
Singapore & 3.11 & 0.99 & 0.19 & 0.11 & 0.08 & 0.18 & 0.18\\
Equatorial Africa & 3.14 & 1.10 & 0.61 & 0.75 & 0.59 & 0.54 & 0.54 \\
East Pacific & 2.94 & 1.08 & 0.22 & 0.12 & 0.09 & 0.22 & 0.21 \\
\bottomrule
\end{tabular}
\end{table}

\subsection{Global data assimilation}
\begin{figure}[h!]
    \centering
    \includegraphics[width=\textwidth]{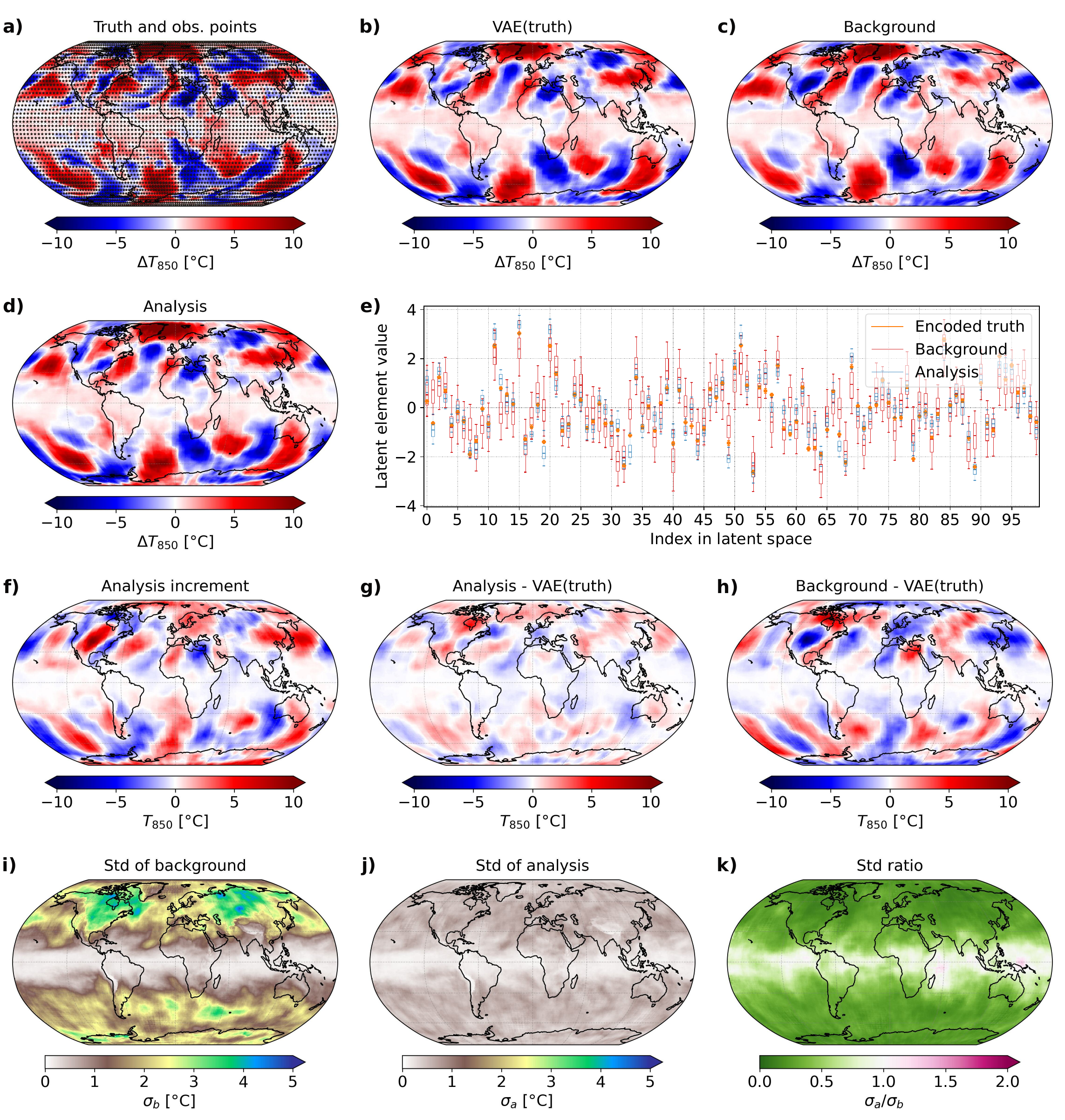}
    \caption{Single cycle data assimilation experiment for April 15, 2019, with 4050 observations with the longitudinal and latitudinal spacing of $4^\circ$ scattered around the globe. The observations were sampled as described in~\ref{sec:osse} and their $\sigma_o$ was set to 1\,K. (a) The ground truth $\mathbf{x}_t^d$ and the locations of the observations. (b) The mean of the VAE-reconstructed truth $D\left(\mu_\phi\left(\mathbf{x}_t^d\right)\right)$, (c) The mean background, represented in the grid point space. (d) As (c), but for the analysis. (e) The distributions in the latent space for the encoded truth (orange), background (red) and analysis (blue). See Figure~\ref{fig:singobs_Ljubljana} for the box and whisker plot properties. (f) The analysis increment (i.e.~panel (d) minus panel (c)). (g) The analysis with subtracted VAE of truth (i.e.~panel (d) minus panel (b)). (h) The background with subtracted VAE of truth (i.e.~panel (c) minus panel (b)). (i) The standard deviation of the background. (j) The standard deviation of the analysis. (k) The ratio between the standard deviations of background and analysis ($<1$ is good).}
    \label{fig:multiple_obs}
\end{figure}

In our final experiment, we analyse the performance of NNDA in the case of many observations, evenly distributed across a global latitude-longitude grid with $4^\circ$ spacing (Figure~\ref{fig:multiple_obs}a). These observations were simulated from the ground truth for day $d$, while the background was based on encoded truth from day $d-1$, as detailed in Section~\ref{sec:osse}. The observation-error standard deviation $\sigma_o$ was constant and set to~1\,K, while the background-error standard deviation is nonhomogeneous and varies between 0.1 K in the Tropics, and 4 K in the middle and high latitudes, as depicted in Figure~\ref{fig:multiple_obs}i. Panels \ref{fig:multiple_obs}b,c depict the reconstructed truth and the background, while panel \ref{fig:multiple_obs}d illustrates the resulting analysis following the assimilation of global observations. As expected, the analysis aligns more closely with the reconstructed truth compared to the background (Figure~\ref{fig:multiple_obs}g,h). The cosine-weighted root-mean-square error (RMSE) of the background, evaluated against the ERA5 ground truth (Figure~\ref{fig:multiple_obs}a) is 2.5\,K, whereas the RMSE of the analysis is 1.9\,K. Similarly, the analysis distribution also shifts towards the mean-encoded truth ($\mu_\phi (\mathbf{x}_t^d)$) for most latent vector elements (Figure~\ref{fig:multiple_obs}e). The RMSE of the background latent state evaluated against the truth latent state is 0.34, whereas the RMSE of the analysis latent state is 0.13. A relatively smaller relative reduction of analysis RMSE compared to background RMSE in the grid point space is mainly a consequence of the limited, coarse-grained representation of the $\teightfifty$ field by our VAE. Moreover, the analysis-error standard deviation is significantly reduced in comparison to the background-error standard deviation, except in the tropics where $\sigma_o \gg \sigma_b$ (Figure~\ref{fig:multiple_obs}i,j,k). This feature is also evident in the latent space, where the analysis distribution of each latent vector element is narrower than the corresponding background distribution (Figure~\ref{fig:multiple_obs}e). The mean value of the standard deviation of the latent state elements is also reduced from 0.66 for the background to 0.27 for the analysis. 

\begin{figure}[h!]
    \centering
    \includegraphics[width=\textwidth]{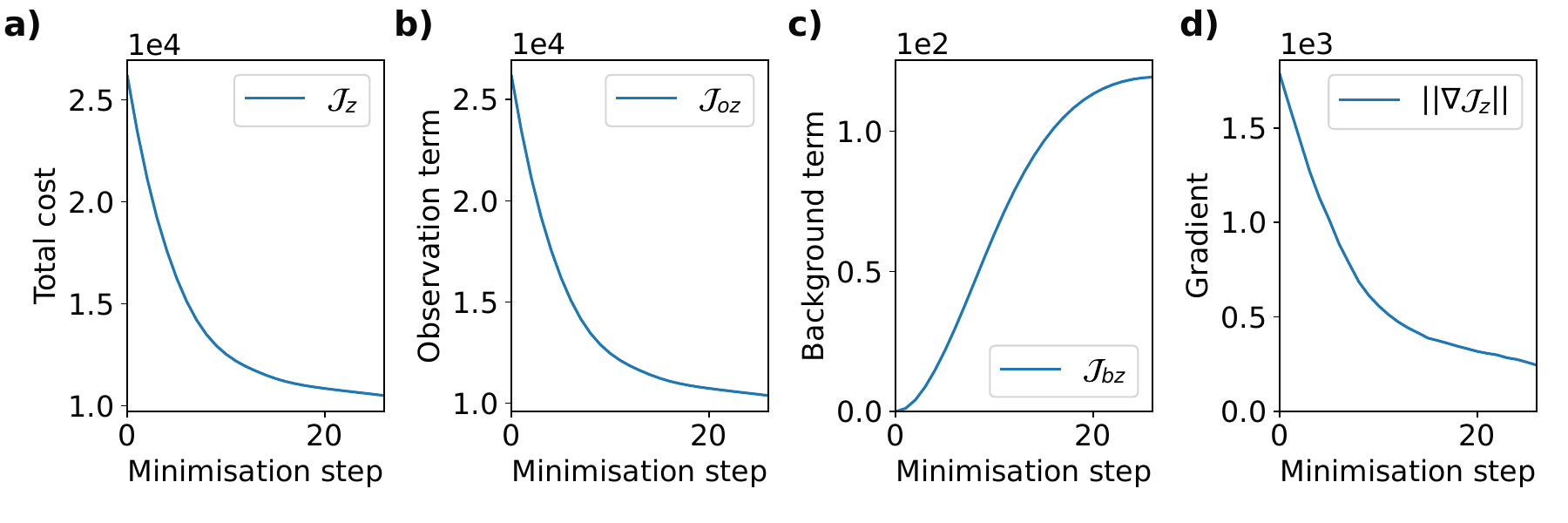}
    \caption{Cost function convergence for a single ensemble member from the global DA experiment (Figure~\ref{fig:multiple_obs}). (a) The total value of the cost function $\mathcal{J}$~(Equation~\ref{eq:3D-Var latent space}), (b) the value of the observation term $\mathcal{J}_o$, (c) the value of the background term $\mathcal{J}_b$, and (d) the Euclidean norm of the cost function gradient $\nabla\mathcal{J}_z$. Note the difference in the order of magnitude between (a,b), (c) and (d).}
    \label{fig:cost_function}
\end{figure}

The convergence of the cost function $\mathcal{J}_z$ (\ref{eq:3D-Var latent space}) during the minimisation process is demonstrated in Figure~\ref{fig:cost_function}. The minimisation stopping criterion (\ref{eq:stopping_criterion}) was achieved after 26 iterations in this case. 

\section{Discussion, Conclusions and Outlook}\label{section:conclusions} 

The primary objective of this study was to present a proof-of-concept for variational data assimilation with neural networks in numerical weather prediction. To achieve this, we trained a convolutional variational autoencoder (VAE) to represent global temperature fields at 850\,hPa pressure level ($\teightfifty$) from ERA5 reanalysis in a reduced-dimension latent space (Figs.~\ref{fig:VAE_structure}, \ref{fig:VAE_of_truth}). This enabled the construction of an auto-differentiable three-dimensional variational (3D-Var) cost function (\ref{eq:3D-Var latent space}) in the latent space, which measures the distance of the latent vector to the observations and the background latent vector. We employed the Adam optimiser, a stochastic gradient descent method, for minimising the cost function to derive the optimal analysis, utilising auto-differentiation of the decoder (Equation~(\ref{eq:cost_function_gradient})). The background-error covariance in the latent space was estimated from the 24-hour differences of encoded ground truth states. We have shown that the $\mathbf{B}_z$-matrix in the neural-network derived space is quasi-diagonal with diagonal elements mostly more than an order of magnitude larger than off-diagonal elements (Fig.~\ref{fig:B matrix classic}). Assuming diagonal $\mathbf{B}_z$-matrix, the background-error term in Equations~(\ref{eq:3D-Var latent space}) and~(\ref{eq:cost_function_gradient}) was significantly simplified. While the resulting $\mathbf{B}_z$-matrix is static in the latent space, both background-error variances and covariances exhibit seasonal variations and dependence on the current latent-state-representation of the atmosphere, commonly referred to as flow-dependent behaviour (Fig.~\ref{fig:background std}).

We conducted a number of data assimilation experiments with a single assimilation cycle (Fig.~\ref{fig:osse}), in which the persistence-based background was corrected by assimilating  $\teightfifty$ observations, simulated from ERA5 ground truth. Every experiment comprised an ensemble of 150 data assimilations, with each assimilation performed with differently perturbed background latent states and observations, resulting in an ensemble of 150 analyses. Single observation experiments revealed the structure of the background-error covariances, showcasing significant differences between the midlatitudes and the tropics. These covariances incorporate the large-scale dynamical features of the atmosphere (Figs.~\ref{fig:singobs_Ljubljana}, \ref{fig:singobs_SW_Indian_Ocean}-\ref{fig:singobs_E_Pacific}). We also verified that the magnitude of the increments resulting from NNDA minimisation aligned with the theoretical expectations (Table~\ref{tab:singobs increments}), confirming the accuracy of our method and its implementation. Finally, we conducted a global data assimilation experiment where we monitored the reduction of the root-mean-square error (RMSE) in the analysis with respect to background in both grid point and latent spaces (Fig.~\ref{fig:multiple_obs}).  We observed that the reduction of RMSE was relatively greater in the latent space compared to the grid-point space. This difference can be ascribed to the limited reconstruction ability of our VAE, comprising of a latent state of only 100 elements.

Our approach represents the first example of variational data assimilation in the neural-network-derived reduced latent space for numerical weather prediction (NWP). While previous applications of neural-network DA in latent spaces were confined to Kalman filter methods  \citep{Amendola2021,Peyron2021}, and reduced-space variational DA relied on linear transformations like PCA or SVD \citep{Robert2005,Chai2007,Cheng2010,Mack2020}, our method employs nonlinear neural-network-derived transformations. For instance, \citet{Peyron2021} highlighted that the reduced space obtained with neural networks can be of much smaller dimension than linear model reduction techniques, resulting in reduced condition number and faster convergence. In most NNDA studies, the authors evaluated the observational component of the cost function ($\mathcal{J}_{zo}$) in the latent space. This requires a multistep transformation from the observation space to the model space and then to the latent space with the encoder network \citep[e.g.][]{Mack2020,Amendola2021}. Learning such reconstruction is not trivial in the operational NWP data assimilation, given the complex relations between observed and prognostic variables, along with time-evolving biases in observation systems \citep{Dee2004varbc}. Furthermore, the dimensionality of the model space (total variable count times number of grid points) greatly surpasses the number of observations by more than an order of magnitude. The transformation from the observation space to the model space is therefore a highly underdetermined problem. Additionally, such transformation disseminates observational information throughout the domain, leading to correlated observation errors. Based on that, we believe that the observation operators, whether physical or derived via neural networks, will remain indispensable until a continuous, spatially and temporally comprehensive observation network is achieved. The results of \citet{Andrychowicz2023} further indicate that relying solely on observations for determining the analysis increments is unfeasible. This underscores the need to combine observations with complete prior representation of the atmospheric state in a mathematically well-defined Bayesian framework (which does not need to be relearned from data within the so-called foundational models).

Our work introduces a novel approach to a unified representation of background-error covariances in the tropics and the midlatitudes. In contrast, the operational NWP DA employs background-error covariance models that capture large-scale atmospheric features characterised by low Rossby numbers, such as extratropical flows, not explicitly accounting for mesoscale and tropical balances \citep{Bannister2021}. The background-error covariance model devised in our study demonstrates enhanced correlations in the south-west to north-east direction in the Northern Hemisphere midlatitudes, in alignment with the prevailing south-westerly winds. Negative correlations at a distance of around 2000 km from the observation site align with the previous studies for the climatological $\mathbf{B}$ \citep[e.g.][]{Fisher2003}. In the tropics, background-error correlations exhibit larger correlation lengthscales in the zonal direction, in accordance with the mean easterlies in the equatorial lower troposphere and disturbances confined within the equatorial waveguide. Following an assimilation of an observation with positive temperature departure, the shapes of the analysis increments resemble the Gill-like response \citep{Gill1980} to the diabatic heating (Figs.~\ref{fig:singobs_Singapore}, \ref{fig:singobs_CAR}). This temperature perturbation also induces the excitation of the Rossby wave trains, in line with the experiments presenting adjustment of global atmospheric flow to heating perturbation in the tropics \citep[e.g.][]{Trenberth1998,Kosovelj2019}. Single observation experiments, such as the one with the temperature observation over the Eastern equatorial Pacific, resulted in increment extending into the central Pacific, resembling the ENSO pattern. At the same time,  the increment did not extend over the Andes into the Amazon, implicitly taking into account the land-sea distribution and orography. While these teleconnecting patterns are physically meaningful, it remains uncertain whether such large-scale increments would benefit the operational NWP DA. 

The impact of long-distance correlations was amplified in our study due to the excessive $\sigma_b$ in the midlatitudes and underestimated $\sigma_b$ in the tropics, a common issue in climatology-based estimation of background-error covariances \citep{Bannister2008}. To fight this issue, we attempted to construct a flow-dependent $\latentB$ from 10-member (9 plus control) ERA5 ensemble. While the ensemble size is insufficient for obtaining adequate background-error variance statistics, and the resolution constraints of our VAE hindered the representation of the finely-structured flow-dependent background-error variances, our additional experiments (not shown) revealed that the VAE typically makes the forecast distribution more Gaussian in the latent space and also produces a quasi-diagonal background-error covariance matrix. 

In this study, we selected VAE over the standard AE to ensure that the latent state vector follows the Gaussian distribution and a smooth transition from the latent to the grid point space. The Gaussian nature of the latent state proves highly advantageous for variational NNDA, similar to how the assumption of gaussianity of the control variables is central in the derivation of variational data assimilation \citep{Lorenc1986}. An important limitation of using VAE instead of standard AE is that most neural-network-based forecast models use standard autoencoders. Employing data assimilation and forecasting model within the same space would reduce the number of encoder and decoder transformations, thereby reducing the complexity of a single data assimilation cycle. One possible solution to this problem is to use a Probabilistic Autoencoder that learns the probability distribution of the AE latent space weights using a normalizing flow \citep{Bohm2020}. 

There are several other limitations of our study that need to be addressed in the future. We employed the simplest possible (persistence-based) forecast model of a single atmospheric variable ($\teightfifty$), no model error was applied, and we simulated observations from ERA5. This setup limited the usability of cycling experiments, which are typically employed to assess the benefits of the newly proposed DA algorithm. Therefore, further improvements of the autoencoder model are needed, which would include more variables, higher horizontal resolution and multiple vertical levels and an extended latent space, before relying on the latent space $\mathbf{B}$-model in operational variational DA. A comparison should be made between the presented neural-network 3D-Var method and the traditional 3D-Var method, both in terms of accuracy and computational gains. Furthermore, the method should be extensively tested in the multivariate setup to test whether it produces reliable cross-covariances of errors between different model fields. Our extra experiments (not shown) revealed that the method is capable of representing multivariate background-error covariances for 200 hPa horizontal wind and geopotential, producing geostrophically-balanced analysis increments in the midlatitudes. Nonetheless, additional experiments with other variables and multiple levels are needed.

Work has started to extend the presented method to resemble 4D-Var by merging the convolutional-autoencoder-based forecast model of \citet{Perkan2023a} and probabilistic autoencoder \citep{Bohm2020}. Auto-differentiable NN or hybrid forecast models are particularly advantageous in 4D-Var data assimilation, where deriving the tangent-linear model and its adjoint demands significant effort \citep{Janiskova2013}.  While NN forecast models particularly excel at short forecast lead times, we aim to explore the prospect of extending the assimilation window. We claim, that the 4D-Var approach is the way to go forward, as it allows to make full use of tracer data through the 4D-Var tracing effect \citep{Zaplotnik2023} and mass data (e.g.~temperature) through the internal geostrophic adjustment process \citep{Zaplotnik2018}. 

The initialisation of the model forecast through data assimilation is the last missing link to performing standalone neural-network weather prediction and generating new reanalysis training data for a new generation of neural network models. In view of the looming climate crisis and substantial energy savings offered by NN prediction and training relative to traditional NWP models, a concerted effort should be made to develop a reliable NNDA as soon as possible.

\section*{Acknowledgements}
Bo\v{s}tjan Melinc is supported by ARRS Programme P1-0188. The authors would like to thank Gregor Skok (UL FMF) for consistent support in acquiring the PhD funding. The authors acknowledge the work of Philip Brohan (UKMO) on Machine Learning for Data assimilation (\url{http://brohan.org/Proxy_20CR/}), who generously shared the initial code for this research (\url{https://github.com/philip-brohan/Proxy_20CR}). The authors are grateful to Massimo Bonavita (ECMWF) for providing thoughtful suggestions on the early manuscript and Mariana Clare (ECMWF) for carefully reading the manuscript and fruitful discussions on the topic.

\section*{Conflict of interest}
The authors declare no conflict of interest.

\bibliographystyle{abbrvnat}
\bibliography{references}

\appendix
\section*{Appendix A: Proof of sufficient ensemble size for representing daily flow-dependent features of background standard deviation}\label{appendix:a}

\begin{figure}[h!]
    \centering
    \includegraphics[width=\textwidth]{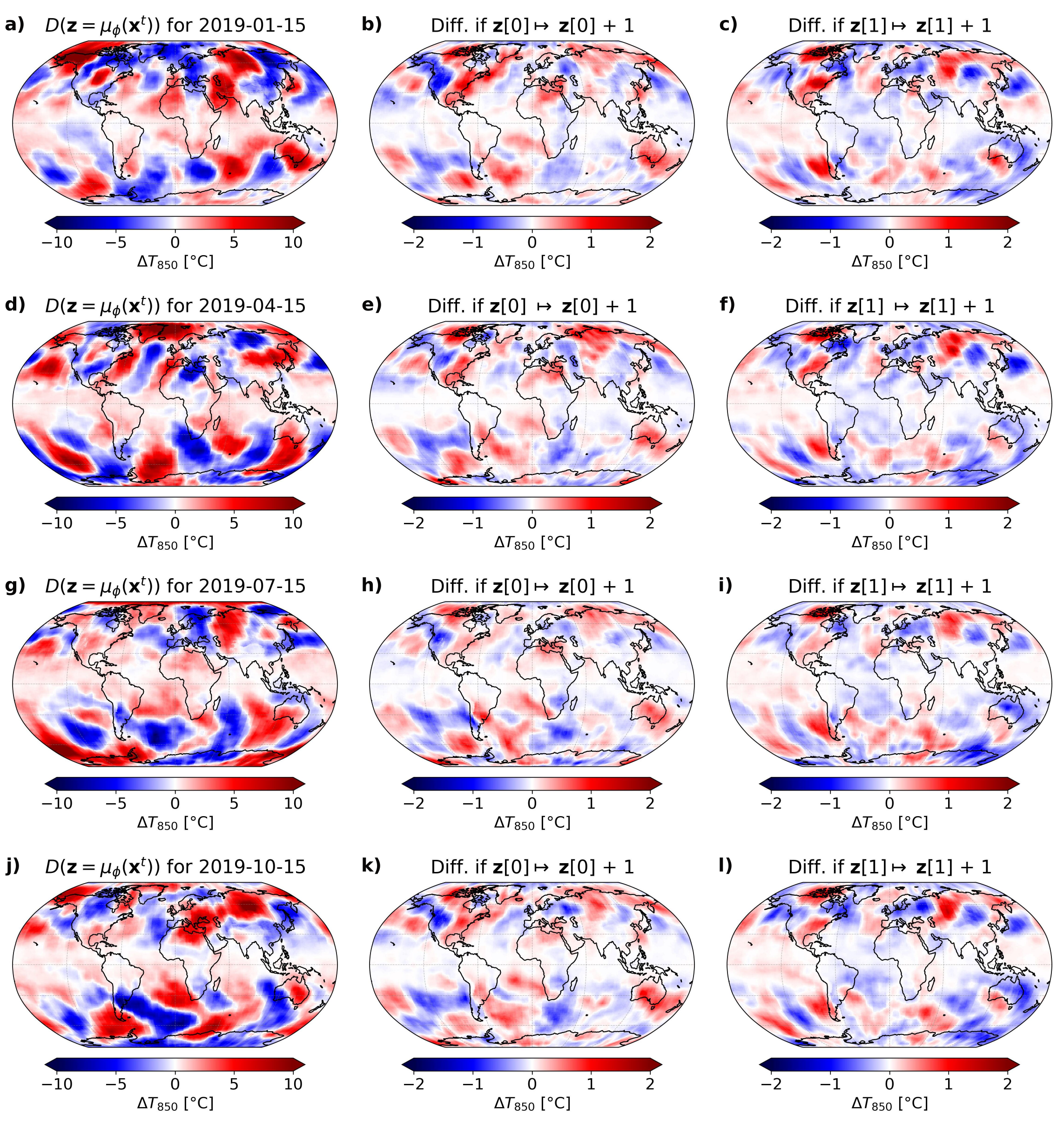}
    \caption{As Figure~\ref{fig:mappings}a,b,d, but for (a-c) January 15, 2019, (d-f) April 15, 2019, (g-i) July 15, 2019, and (j-l) October 15, 2019.}
    \label{fig:mappings4}
\end{figure}

\begin{figure}[h!]
    \centering
    \includegraphics[width=\textwidth]{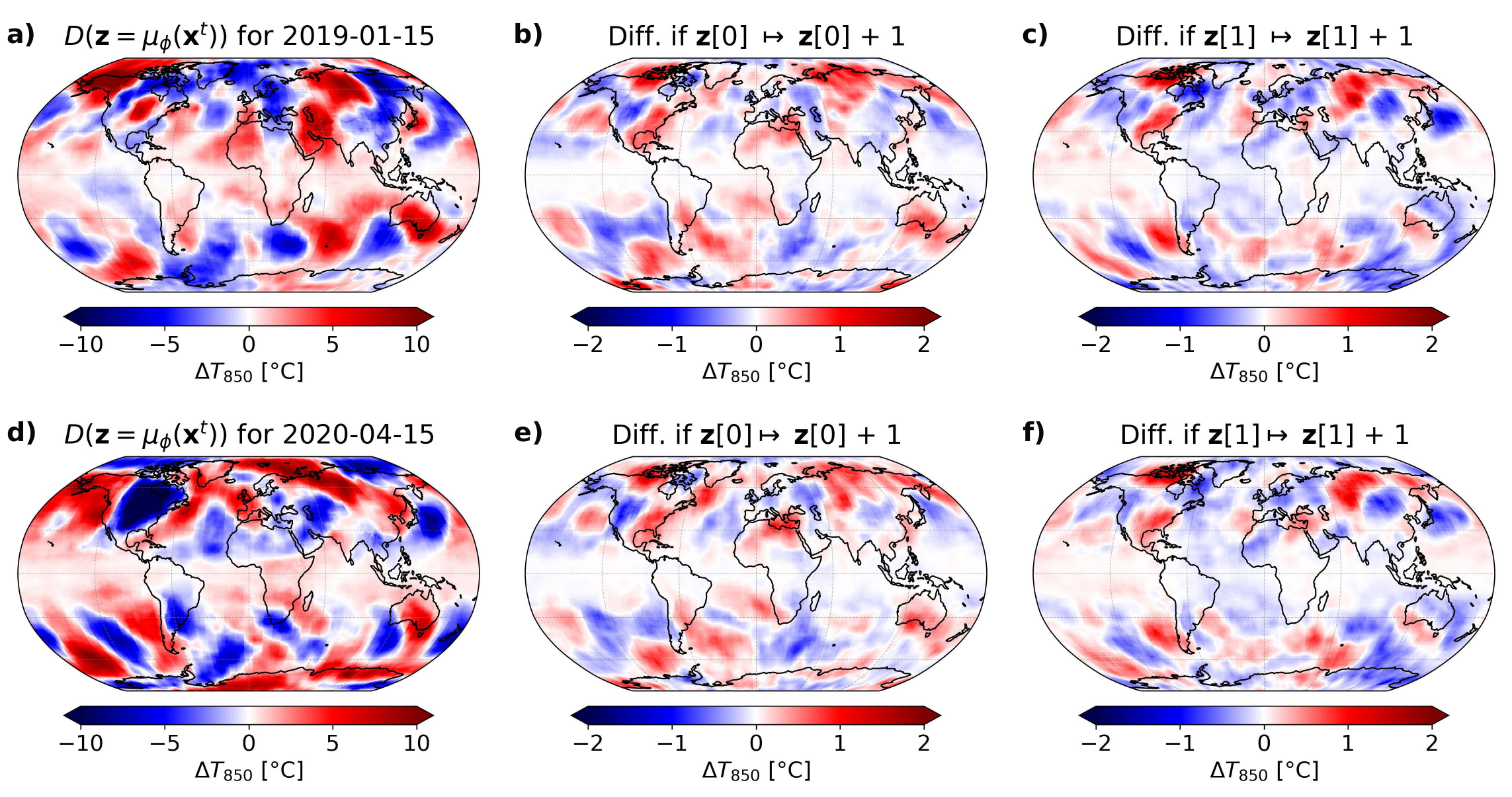}
    \caption{(a-c) Reproduced Figure~\ref{fig:mappings}a,b,d. (d-f) As (a-c), but for April 15, 2020.}
    \label{fig:mappings2}
\end{figure}

In Section~\ref{section:mappings}, we presented mappings from the latent space to the grid-point space, corresponding to specific latent vector elements for April 15, 2019, noting their date-specific variability. Figure~\ref{fig:mappings4} illustrates consistent large-scale patterns across seasons, varying mainly in magnitude due to seasonal shifts in climatological standard deviations. However, the differences in the smaller scale features below approximately $\SI{1000}{\kilo\meter}$ are clearly visible.
The mappings do not only vary due to destandardisation of the decoder's output, which is day-of-the-year dependent, but also due to the nonlinearity of decoder. Figure~\ref{fig:mappings2} displays mappings for two distinct global weather situations on the same day-of-year but different year, highlighting the differences between the two. This can be explained as follows. Lets assume that the same perturbation to the first latent vector element $\delta\mathbf{z}=[1,0\ldots,0]^T$ is added to different original latent states $\mathbf{z}_1$ and $\mathbf{z}_2$ that represent two distinct atmospheric states. The perturbed latent states read $\mathbf{z}_{p1}=\mathbf{z}_1+\delta\mathbf{z}$ and $\mathbf{z}_{p2}=\mathbf{z}_1+\delta\mathbf{z}$. The grid point perturbation corresponding to the perturbation of first latent vector element is then
\begin{equation}\label{eq:pert1}
\begin{aligned}
    \delta\mathbf{x}_1  &= \mathbf{x}_{p1} - \mathbf{x}_1 = \\
                        &= D(\mathbf{z}_{p1}) - D(\mathbf{z}_{1}) = \\
                        &\approx D(\mathbf{z}_{1}) + \frac{\partial D}{\partial\mathbf{z}}\Bigr|_{\mathbf{z}=\mathbf{z}_1} \delta\mathbf{z} - D(\mathbf{z}_{1}) = \\
                        &= \frac{\partial D}{\partial\mathbf{z}}\Bigr|_{\mathbf{z}=\mathbf{z}_1} \delta\mathbf{z} \, ,
\end{aligned}    
\end{equation}
where we applied first-order Taylor approximation of the decoder $D(\mathbf{z}_{p1})$ around the unperturbed state $\mathbf{z}_{1}$. Term $(\partial D/\partial\mathbf{z})_{\mathbf{z}=\mathbf{z}_1}$ is the linearised decoder, evaluated at $\mathbf{z}_1$. Analogously, we can define 
\begin{equation}\label{eq:pert2}
    \delta\mathbf{x}_2 \approx \frac{\partial D}{\partial\mathbf{z}}\Bigr|_{\mathbf{z}=\mathbf{z}_2} \delta\mathbf{z} .
\end{equation}
We have thus shown that the perturbations $\delta\mathbf{x}_1$ and $\delta\mathbf{x}_2$ in grid point space differ in case of the same perturbation $\delta\mathbf{z}$ applied to distinct latent vectors. As a consequence, the constant background error standard deviations in the latent space ($\mathbf{B}_z$ is constant) produce variable latent-state-dependent background error standard deviations $\sigma_b$ in grid point space, illustrated in Figure~\ref{fig:background std}.

Although we have theoretically demonstrated that the background-error standard deviations differ (Figure~\ref{fig:background std}a,b) as a function of the background latent vector, it could be contended that these differences are modest and primarily discernible due to the finite 150-member ensemble.

\begin{figure}[h!]
    \centering
    \includegraphics[width=\textwidth]{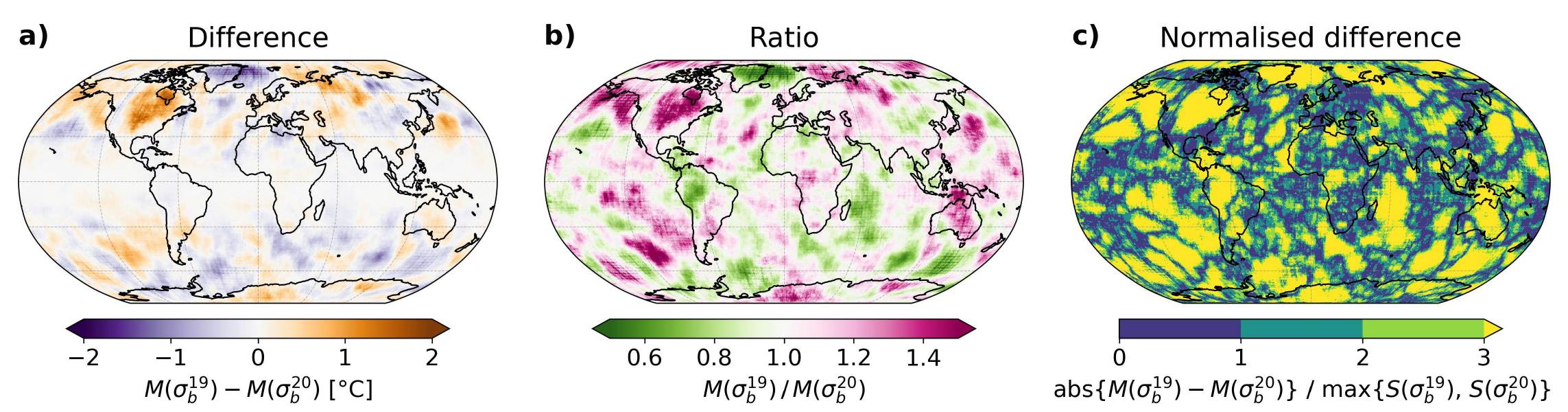}
    \caption{(a) The difference between $M(\sigma_b^{19})$ and $M(\sigma_b^{20})$ in\,K. (b) The ratio between the same quantities. (c) The difference between $M(\sigma_b^{19})$ and $M(\sigma_b^{20})$, normalised with the standard deviation.}
    \label{fig:ensemble_size}
\end{figure}

To assess whether the 150-member ensemble is sufficient to capture the flow-dependent features of $\sigma_b$, we generated 100 such 150-member ensembles for April 15, 2019 and April 15, 2020. The same day of the year was used to eliminate the day-of-year dependent features. Let $\sigma_b^{19}$ denote the background standard deviation of a single ensemble for April 15, 2019, and $\sigma_b^{20}$ the background standard deviation of a single ensemble for April 15, 2020. Further, let $M(\sigma_b^{19})$ be the mean of $\sigma_b^{19}$ over all 100 ensembles and $M(\sigma_b^{20})$ be the analogous quantity for $\sigma_b^{20}$. Finally, let $S(\sigma_b^{19})$ be the standard deviation of $\sigma_b^{19}$ over all 100 ensembles and, again, $S(\sigma_b^{20})$ be the analogous quantity for $\sigma_b^{20}$.
Figures~\ref{fig:ensemble_size}a and~\ref{fig:ensemble_size}b show the difference between $M(\sigma_b^{19})$ and $M(\sigma_b^{20})$ and their ratio. A sufficient measure $m$ for the unbiased confirmation that two datasets can be distinguished one from another is to compare the difference of their means with their standard deviations. Larger distances in terms of standard deviation lead to greater confidence that the datasets are distinct. Consequently, to consider the worst case scenario, one should always divide the difference of the means with the largest of both standard deviations, i.e.
\begin{equation}
    m = \frac{\left|M(\sigma_b^{19}) - M(\sigma_b^{20})\right|}{\mathrm{max}\left(S(\sigma_b^{19}),S(\sigma_b^{20})\right)}
\end{equation}
Figure~\ref{fig:ensemble_size}c shows this metric for $\sigma_b^{19}$ and $\sigma_b^{20}$. We found that for $66\%$ of the Earth's surface, $M(\sigma_b^{19})$ and $M(\sigma_b^{20})$ are at least 1 standard deviation apart, for $39\%$ of the Earth's surface they are at least 2 standard deviations apart and for $22\%$ of the Earth's surface the are at least 3 standard deviations apart. Thus we are confident that the 150-member ensemble is sufficiently large to capture the situation dependent features of $\sigma_b$.

\end{document}